\documentclass[preprint,12pt]{elsarticle}

\usepackage{amssymb}
\usepackage{amsmath,amsfonts}
\usepackage{algorithmic}
\usepackage{algorithm}
\usepackage{array}
\usepackage{textcomp}
\usepackage{stfloats}
\usepackage{url}
\usepackage{verbatim}
\usepackage{graphicx}
\usepackage{CJK}
\usepackage{indentfirst}
\usepackage{cases}
\usepackage{amsfonts}
\usepackage{amssymb,amsmath,amscd}
\usepackage{multirow}
\newtheorem{thm}{Theorem}
\newtheorem{lem}{Lemma}
\newtheorem{Definition}{Definition}

\newtheorem{remark}{Remark}
\newtheorem{prop}{Proposition}

\newtheorem{example}{Example}

\newtheorem{experiment}{Experiment}
\newproof{pf}{Proof}

\journal{Nuclear Physics B}

\begin{document}

\begin{frontmatter}

%% Title, authors and addresses

%% use the tnoteref command within \title for footnotes;
%% use the tnotetext command for theassociated footnote;
%% use the fnref command within \author or \address for footnotes;
%% use the fntext command for theassociated footnote;
%% use the corref command within \author for corresponding author footnotes;
%% use the cortext command for theassociated footnote;
%% use the ead command for the email address,
%% and the form \ead[url] for the home page:
%% \title{Title\tnoteref{label1}}
%% \tnotetext[label1]{}
%% \author{Name\corref{cor1}\fnref{label2}}
%% \ead{email address}
%% \ead[url]{home page}
%% \fntext[label2]{}
%% \cortext[cor1]{}
%% \affiliation{organization={},
%%             addressline={},
%%             city={},
%%             postcode={},
%%             state={},
%%             country={}}
%% \fntext[label3]{}

\title{Frequency Analysis with Multiple Kernels and Complete Dictionary}

%% use optional labels to link authors explicitly to addresses:
 \author[label1]{Cuiyun Lin}
 \affiliation[label1]{organization={Faculty of Innovation Engineering
School of Computer Science and Engineering, Macau University of Science and Technology},
             addressline={Taipa},
             city={Macau},
             postcode={999078},
             %state={},
             country={China}}

 \author[label2]{Tao Qian}
 \affiliation[label2]{organization={Macau Center for Mathematical Sciences, Macau University of Science and Technology},
             addressline={Taipa},
             city={Macau},
             postcode={999078},
             %state={},
             country={China}}

\begin{abstract}
%% Text of abstract
In signal analysis, among the effort of seeking for efficient representations of a signal into the basic ones of meaningful frequencies, to extract principal frequency components, consecutively one after another or $n$ at one time, is a fundamental strategy.  For this goal, we define the concept of mean-frequency and develop the related frequency decomposition with the complete Szeg\"o kernel dictionary, the latter consisting of the multiple kernels, being defined as the parameter-derivatives of the Szeg\"o kernels. Several major energy matching pursuit type sparse representations, including greedy algorithm (GA), orthogonal greedy algorithm (OGA), adaptive Fourier decomposition (AFD), pre-orthogonal adaptive Fourier decomposition (POAFD), $n$-Best approximation and unwinding Blaschke expansion, are analyzed and compared. Of which an order in re-construction efficiency between the mentioned algorithms is given based on  detailed study of their respective remainders. The study spells out the natural connections between the multiple kernels and the related Laguerre system, and in particular shows that both, like the Fourier series, extract out the $O(n^{-\sigma})$ order convergence rate from the functions in the Hardy-Sobolev space of order $\sigma >0.$ Existence of the $n$-Best approximation with the complete Szeg\"o dictionary is proved and the related algorithm aspects are discussed. The included experiments form a significant integration part of the study, for they not only illustrate the theoretical results, but also provide cross comparison between various ways of combination between the matching pursuit algorithms and the dictionaries in use. Experiments show that the complete dictionary remarkably improves approximation efficiency.
\end{abstract}

%%Graphical abstract
\begin{graphicalabstract}
\end{graphicalabstract}

%%Research highlights
\begin{highlights}
\item Research highlight 1
\item Research highlight 2
\end{highlights}

\begin{keyword}
%% keywords here, in the form: keyword \sep keyword

%% PACS codes here, in the form: \PACS code \sep code

%% MSC codes here, in the form: \MSC code \sep code
%% or \MSC[2008] code \sep code (2000 is the default)
complete Szeg\"o kernel dictionary, multiple kernels, pre-orthogonal adaptive Fourier decomposition, $n$-Best approximation, unwinding Blaschke expansion, Laguerre system.
\end{keyword}

\end{frontmatter}

%% \linenumbers

%% main text
\section{Introduction}\label{section:1}
\label{}

%% The Appendices part is started with the command \appendix;
%% appendix sections are then done as normal sections
%% \appendix

%% \section{}
%% \label{}

%% If you have bibdatabase file and want bibtex to generate the
%% bibitems, please use
%%
%%  \bibliographystyle{elsarticle-num}
%%  \bibliography{<your bibdatabase>}

%% else use the following coding to input the bibitems directly in the
%% TeX file.
\def\H{\mathcal H}
\def\D{\mathcal D}
\def\E{\mathcal E}
\def\bD{\bf D}
\def\bC{\bf C}
\def\bR{\bf R}
\def\K{\mathcal K}
Matching pursuit, as a methodology to generate sparse representations of signals, is usually based on a dictionary of the underlying Hilbert space. The most basic matching pursuit algorithm would be one called \emph{greedy algorithm} in the context of Hilbert space with a dictionary \cite{MZ,Temlyakov}. In 2011, Qian and Wang proposed \emph{AFD}, or \emph{Core AFD} \cite{QW,CQT,Qian1}, which crucially uses energy matching pursuit, as well as the complex Hardy space techniques, to develop a sparse representation in the form of a Takenaka-Malmquist system. Ever since then, there have been generalizations and variations of AFD, including \emph{pre-orthogonal AFD} \cite{Q2D}, the $n$-\emph{Best AFD }\cite{WQ3} and \emph{Unwinding AFD} \cite{Qian1+1}. A closely related method, called unwinding Blaschke expansion was proposed early by R. Coifman et al. in 2000 \cite{CS,CP}. AFD was further extended to multivariate and matrix-valued cases \cite{Q2D,ACQS1,ACQS2}. In this paper, we restrict ourselves only to the one dimensional cases with scalar function values. AFD and its one dimensional variations are all based on the Szeg\"o kernel dictionary. We, however, adopt a general formulation as follows.

Let $\H$ be a Hilbert space with a dictionary $\D.$   That means that $\D$ consists of norm-one elements and ${\rm span}\{\D\}$  is a dense subset of $\H.$ Each of the elements of the dictionary $\D$ is labeled by a parameter $q\in \E,$ and is denoted as $E_q.$ We will call a set ${\K}$ of elements in $\H$ a \emph{pre-dictionary} if the unimodular normalizations of the elements in  ${\K}$ form a dictionary. In a reproducing kernel Hilbert space (RKHS), for instance, $K_q(p)\triangleq K(q,p)$ form a pre-dictionary. The parameter set $\E$ is usually an open set of Euclidean space. We usually assume that the concerned $E_q$ and $K_q$ are smooth in $q,$ and in particular have as many orders of derivatives in $\E$ as we use. The boundary of $\E$ in the Euclidean topology is denoted $\partial \E.$ The concerned general theory will be applicable to RKHS and many non-RKHS cases as well.

In the present paper, we will study the case where $\H$ is the complex Hardy space
\begin{align}
 H^2({\bD})&=\{f: {\bD}\to {\bC}\ : f\ {\rm is\ analytic\ in}\  {\bD}, \  \|f\|^2_{H^2({\bD})} \nonumber\\ &=\sup_{0<r<1}\frac{1}{2\pi}\int_0^{2\pi}|f(re^{it})|^2dt<\infty\}
\end{align}
together with the Szeg\"o dictionary,
   \begin{equation}\label{Szego} {\D}=\{e_a\}_{a\in {\bD}}, \quad e_a(z)=\frac{\sqrt{1-|a|^2}}{1-\overline{a}z}, \quad a, z\in {\bD}.\end{equation}
It is noted that for every $f\in H^2({\bD})$
\[ \lim_{{\bD}\ni z\to e^{it}}f(z) \quad {\rm exist\ for\ a.e.}\quad t\in [0,2\pi],\]
where the limit $z\to e^{it}$ takes the non-tangential manner.
The mapping that sends $f(z)\in H^2({\bD})$ to its non-tangential boundary limit $f(e^{it})$ on the unit circle is an isometric isomorphism between $H^2({\bD})$ and a closed subspace $H^2(\partial {\bD})$ of $L^2(\partial {\bD}).$ In terms of the non-tangential boundary limit functions the inner product of $H^2({\bD})$ may be defined through that of the $L^2(\partial {\bD})$ space, namely,
\begin{eqnarray}\label{given} \langle f,g\rangle_{H^2}=\frac{1}{2\pi} \int_0^{2\pi}
f(e^{it})\overline{g}(e^{it})dt.\end{eqnarray} Using this inner product, as a consequence of the Cauchy formula, $H^2({\bD})$ is a RKHS having $k_a(z)$ as its reproducing kernel, where
\[ k_a(z)=\frac{1}{1-\overline{a}z}.\] The Szeg\"o dictionary element $e_a$ is the norm-1 normalization of the Szeg\"o kernel $k_a.$ The collection ${\mathcal K}=\{k_a\}_{a\in {\bD}}$ is a pre-dictionary.

The space of the boundary limit functions can be alternatively defined as
\begin{align}
&H^2(\partial {\bD})\nonumber\\
&=\{f: {\partial \bD}\to {\bC}\ : \ f(e^{it})=\sum_{k=0}^\infty c_ke^{ikt}, \nonumber\\
&\|f\|^2_{H^2({\partial \bD})}\triangleq \sum_{k=0}^\infty |c_k|^2<\infty \}.
\end{align}

\begin{remark}
For any real-valued function of finite energy $g\in L^2(\partial {\bD}),$
\[ g^+\triangleq\frac{1}{2}(g+iHg+c_0)\in H^2({\bD}),\]
\[ g^-\triangleq\frac{1}{2}(g-iHg-c_0)\in H^2(\overline{\bD}^c),\]
the latter being the complex Hardy space outside the closure of the unit disc, and
\begin{eqnarray}\label{parti} g=g^++g^-= 2{\rm Re}g^+-c_0,\end{eqnarray}
where $c_0$ is the average of $g$ over the unit circle:
\[ c_0=\frac{1}{2\pi}\int_0^{2\pi}g(e^{it})dt,\]
$H$ is the circular Hilbert transform. For proofs of these fundamental relations, see, for instance, \cite{Garnett} or \cite{Rudin}. The relation (\ref{parti}), in particular, reduces the analysis of the real-valued functions on compact intervals to that of the functions in the Hardy space. The latter is, in fact, the space of the $Z$-transforms.

There exists a parallel theory for functions $g$ defined on the real line with finite energy in which $g^\pm\triangleq\frac{1}{2}(g+iHg),$  where
$H$ is the Hilbert transform on the line and $g=2{\rm Re}g^+.$ $g^+$ is identical to the Laplace transform of $g.$ In both cases, $g^+$ is interpreted as the Gabor analytic signal associated with $g.$ In the real line case, the associated Hardy space is $H^2({\bC}^+)$ that is isometric with
$H^2(\partial {\bC}^+)=H^2({\bR}).$

The same philosophy is obeyed by signals of several complex and several real variables (with the Clifford algebra setting), and of scalar- or vector-, and even matrix-values. Some of the cases have been developed.  See \cite{Q2D, ACQS1, ACQS2, QSW} and the references therein.
\end{remark}

In the rest of the paper we concentrate on developing the theory and practice in the unit disc case, and when we use the notation $H^2$ and the terminology $\lq\lq$the Hardy space" we refer to just the unit disc case.

With the matching pursuit idea the following \emph{ $n$-Best} question is natural.\\

\noindent{\bf The Ill-Posed $n$-Best Question:}

Let $f\in \H$ and $n$ be a fixed positive integer.   Can one find $n$ distinguished parameters $q_1,\cdots,q_n$ and $n$ complex numbers $c_1,\cdots,c_n$ such that for
the corresponding pre-dictionary elements $K_{q_1},\cdots,K_{q_n},$ there holds
\label{dis}
\begin{align}\label{dis}
\|f-\sum_{k=1}^n c_k {K}_{q_k}\|_{\H}=\inf \{\|f-\sum_{k=1}^n d_k {K}_{p_k}\|_{\H}\ :\ \nonumber \\{\rm all\ distinguished}\ p_k\in \E \ {\rm and\ all}\ d_k\in {\bC}\ \}.
\end{align}
The answer to this question is dependent on the underlying Hilbert space $\H$ and the pre-dictionary in use. Even for $n=1$ the answer is not necessarily $\lq\lq$Yes". See, for instance, \cite{QD2}, for some counterexamples as weighted Hardy type spaces.
\begin{Definition}
\emph{In the study of matching pursuit the concept boundary vanishing condition (BVC) is established: A pair $(\H,\D)$ is said to satisfy BVC
if for every $f\in \H$ there holds
\begin{equation}\label{BVC} \lim_{q\to \partial \E} \langle f,E_q\rangle =0.\end{equation}}
\end{Definition}
If $(\H,\D)$ satisfies BVC, then through a Bolzano-Weierstrass type compact argument  the above $n$-Best problem has a solution for $n=1.$
It is proved in \cite{QW} that the Hardy space $H^2$ and the Szeg\"o dictionary, as a pair, satisfies BVC, and hence there exist 1-Best solutions in the case. Verification of BVC usually involves detailed analysis.  In general, if $(\H,\D)$ satisfies BVC, then the powerful POAFD matching pursuit is available. See \S \ref{section:3} below for details.

When $n>1$ the answer to the $n$-Best question is $\lq\lq$No" even for the Hardy space and the Szeg\"o dictionary case. The following is an example of the ill-posed-ness for $n>1.$

\begin{example}\label{counter}
We take $f(z)=\frac{1}{(2-z)^2},$ which is a function in the Hardy space $H^2({\bD}),$ and take $n=2.$ The function can be infinitely approximated by linear combinations of two distinguished Szeg\"o kernels. Or, the infimum error is zero. However, any two-term approximation cannot get the zero error.
\end{example}

However, if in the $n$-Best question the phrase
$\lq\lq$all distinguished $p_k\in \E$"  is modified to be $\lq\lq$all \emph{multiple kernels} $\tilde{K}_{p_k},  p_k\in \E,$", then the problem in many cases becomes well-posed. In the Hardy space and Szeg\"o dictionary case, for instance, multiple kernels are defined to be derivatives of $K_a$ with respect to $\overline{a}.$
\def\bN{\bf N}
\def\K{\mathcal K}
\def\tK{\tilde{\mathcal K}}
\def\tD{\tilde{\mathcal D}}

We now introduce multiple Szeg\"o kernels and the complete Szeg\"o kernel dictionary as follows \cite{QianBook}. Denote the set of non-negative integers by $\bN.$
Generating the notations in (\ref{Szego}), we have
\begin{Definition} The set of multiple Szeg\"o kernels
\begin{equation}
\tK=\{k_{n,a}\}_{n\in {\bN},a\in {\bD}},\nonumber\end{equation}
is a pre-dictionary consisting of
\begin{equation}\label{kna}
\quad k_{n,a}(z)=\left(\frac{\partial}{\partial {\bar{a}}}\right)^{n}k_{a}(z)=C_{n,a}\frac{z^n}{(1-\overline{a}z)^{n+1}},\end{equation}
where $C_{n,a}$ are constants depending on $n$ and $a.$
The totality of their unimodular normalizations
\begin{equation}\label{ena} \tD=\{e_{n,a}(z)=\frac{k_{n,a}}{\|k_{n,a}\|}\}_{n\in {\bN},a\in {\bD}}\end{equation} is called the \emph{complete Szeg\"o dictionary} with the parameter set $$\E=\{(n,a)\}_{(n,a)\in {\bN}\times{\bD}}.$$
\end{Definition}

Returning to Example \ref{counter}, $f(z)=1/(z-2)^2$ may be expressed as
$C\frac{\partial}{\partial \overline{a}}k_a,$ where $a=1/2$ and $C$ is a complex number.  The $2$-Best approximation (with infimum error $0$) is then reached.

The multiple kernel notion may be extended to general Hilbert spaces  $\H$ with a pre-dictionary $\D=\{K_q\}, q\in \E.$  To simplify the terminology we will call all the subjects $K_q, E_q, e_a, k_a$ etc., by \emph{kernel}, although some are normalized and some are not, and we will normally denote
 \begin{equation} \{E_q\}=\{\frac{K_q}{\|K_q\|}\}, \quad q\in \E.\end{equation}
Note that for a kernel $K_q(x)$ the domain $\E$ for the parameter $q$ and that for the spatial or the time variable $x$ may not be the same. Let $l$ be a positive integer, and $(q_1,\cdots,q_l)$ an $l$-tuple of parameters in $\E,$ allowing multiplicity. Define $l(q_1,\cdots,q_l)$ be the repeating number of $q_l$ in $q_1,\cdots,q_l, 1\leq l\leq n.$ Without ambiguity, we write $l(q_1,\cdots,q_l)$ as $l(q_l)$ in short. As examples, $l(q_1)=1,$ and $l(q_l)=1$ if $q_l$ is different from the proceeding $q_1,\cdots,q_{l-1}.$
For $\E$ being an open set of the complex number field, we define
\begin{equation}
 \tilde{K}_{q_l}=\bigg[\bigg(\frac{\partial}{\partial \overline q}\bigg)^{(l(q_l)-1)}K_{q}\bigg](q_{l})
\end{equation}
to be the $l$-\emph{th multiple kernels with respect to} $(q_1,\cdots,q_l),$ and $\tilde{E}_{q_l}$ the norm-1 normalization of $\tilde{K}_{q_l}.$ The differential operation $\frac{\partial}{\partial \overline q}$ in below will also be denoted as $\overline{\partial}.$ If $q_l$ are several or hyper-complex variables, then the derivatives are replaced by directional derivatives. In the RKHS case there exists the following useful relation:
\begin{equation}\label{view} \overline{\partial}^{l(q_l)-1}f_{q_k}=\overline{\partial}^{l(q_l)-1}\langle f,K_{q_k}\rangle=\langle f,\tilde{K}_{q_k}\rangle.\end{equation}

Under the multiple kernel concept, the question (\ref{dis}) may be re-formulated to become well-posed:\\

\noindent{\bf The $n$-Best Question (reformulation):}
Let $f\in \H, \K$ a pre-dictionary,  and $n$ a fixed positive integer.   Can one find $n$ parameters $q_1,\cdots,q_n,$ with multiplicity when necessary, and $n$ complex numbers $c_1,\cdots,c_n,$ such that for
the multiple kernels $\tilde{K}_{q_1},\cdots,\tilde{K}_{q_n}$ there holds
\label{diss}
\begin{align}\label{diss}
\|f-\sum_{k=1}^n c_k \tilde{K}_{q_k}\|_{\H}=\inf \{\|f-\sum_{k=1}^n d_k {K}_{p_k}\|_{\H}\ :\ \nonumber \\{\rm all\ distinguished}\ p_k\in \E \ {\rm and\ all}\ d_k\in {\bC}\ \},
\end{align}
where $\tilde{K}_{q_k}, k=1,\cdots,n,$ are, consecutively, the multiple kernels associated with $(q_1, \cdots, q_n).$

Under the new formulation, existence of solution to the $n$-Best problem has been proved for a number of most commonly studied Hilbert spaces of a BVC dictionary, including the Hardy space, the Bergman space, and the weighted Bergman spaces together with the dictionaries naturally induced by their respective reproducing kernels. See \cite{WQ3,QQLZ} and the references therein. The $n$-Best approximation problem in the Hardy space is, in fact, equivalent to the best approximation problem by rational functions in the space of degrees not exceeding $n$ \cite{Walsh}. Several practical algorithms for the Hardy $n$-Best have been proposed that, however, cannot prevent from sinking into the local minima \cite{Bara,Bara2,Qian2,QWM}. Through generalizing the techniques in relation to the backward shift operator, existence of the $n$-Best approximation was lately extended to a class of RKHSs, including the weighted Bergman and weighted Hardy spaces as particular cases \cite{Qian-n-best}. These existence proofs also play a definitive role in seeking for a theoretical algorithm to obtain all the $n$-tuple minimizers. In the present paper, we prove existence of the $n$-Best approximation for the Hardy space under the complete Szeg\"o kernel dictionary. See \S \ref{section:6}.

Besides the one for the n-Best approximation there is another reason that motivates the study of multiple kernels: We are to decompose a signal into its principal frequency components in terms of the energy, not in the degree of the frequency. In \S \ref{section:2} we define the notion mean-frequency for functions in the Hardy space. Mean-frequency is a measurement of the total amount of the frequencies in an analytic signal by which Szeg\"o kernels possess zero mean-frequency. All the concerned matching pursuit algorithms in the context are to select one after another dictionary elements, but unfortunately restricted to only the zero mean-frequency ones. In such a way, the high frequency terms are generated by the GS orthogonalization process, in which the order of applying GS process is a matter. What is desirable in frequency decomposition, however, would be the principal components directly related to the signal. They should  be selected in terms of the greatest energy matching to the complete dictionary elements possessing any frequency, or,  in other words, not be restricted to prescribed frequency levels. This question was also raised by \cite{Borowicz} who pointed out that the existing analytic frequency decomposition, I.e.,  adaptive Fourier decomposition is not according to principal frequency components, but constructed from representatives of the lowest frequency. The present study points out how by using the complete dictionary and performing POAFD the goal of the principal frequency component may be achieved.

The writing plan of this paper is as follows. In \S \ref{section:2} we establish the notion of \emph{mean-frequency} and prove some basic results. Mean-frequency is used to measure the total amount, or degree level, of frequencies that an analytic signal contains. In \S \ref{section:3} we give a concise summary but detailed analysis of the most commonly concerned matching pursuit algorithms. The analyzed algorithms include AFD, GA, OGA, POAFD and $n$-Best. We establish an order between them in accordance with their re-constructing efficiencies. \S \ref{section:4} is devoted to a detailed study of POAFD over the complete Szeg\"o dictionary. In \S \ref{section:5} we study convergence rates in relation to the multiple Szeg\"o kernels and the Laguerre systems. The main results include that, as a generalized form of the Riemann-Lebesgue Lemma, $\langle f,e_{k,a}\rangle $ tends to zero with the order $O(n^{-\sigma})$ for functions $f$ in the Hardy-Sobolev space of order $\sigma>0;$ and as for the Fourier series, the Fourier-Laguerre series in the $\sigma$-Hardy-Sobolev space is of the same convergence rate $O(n^{-\sigma})$. In \S \ref{section:6} we prove existence of the $n$-Best approximation with the complete Szeg\"o kernel dictionary. The existence cannot be deducted from the existing results for RKHSs. We also discuss theoretical and practical algorithms of the $n$-Best solutions. \S \ref{section:7} contains a great number of experiments for comparison between the re-construction efficiencies of AFD, GA, OGA, POAFD, and $n$-Best over, respectively, the Szeg\"o and the complete Szeg\"o dictionaries, as well as with the Unwinding Blaschke expansion. They stand as a significant integration part of the paper. The experiments may be divided into two types, of which one is to verify the theoretical ordering of strongness of the concerned matching pursuit algorithms; and the other is to show strongness of the complete dictionary itself:  The complete Szeg\"o dictionary with weaker algorithms may be stronger than the Szeg\"o dictionary used with stronger type algorithms. In \S \ref{section:8} conclusions and comments are drawn.
\section{Mean-Frequency of signals in the Hardy space}\label{section:2}
It is basic knowledge that any Hardy space function $f\in H^2({\bD})$ has a factorization
$f(z)=\phi(z)s(z)o(z),$ where $\phi (z), s(z)$ and $o(z)$ are, respectively, the Blaschke product, the singular inner function, and the outer function parts of $f$ \cite{Garnett}. The factorization is unique up to unimodular constants. The non-tangential boundary limits of the three functions have, respectively, the forms (in almost everywhere sense on the boundary)
\begin{equation} \phi(e^{it})=e^{i\theta_\phi(t)},\quad s(e^{it})=e^{i\theta_s(t)}, \quad o(e^{it})=\rho_o(t)e^{i\theta_o(t)},\end{equation}
where $\rho_o(t)\ge 0,$ and $\theta_\phi(t), \theta_s(t)$ and $\theta_o(t)$ are real-valued.  As proved in \cite{Qianphasederivative},
\begin{equation}\theta'_\phi\ge 0, \quad  \theta'_s\ge 0,\quad  {\rm a.e.},\end{equation} and
\begin{equation}\int_0^{2\pi} \theta'_\phi(t)dt=N\ge 0,\end{equation}
where $N$ is the number of zeros (can be zero or $\infty$) of the Blaschke product $\phi.$
Moreover, with mild conditions on $f$ to guarantee absolute continuity of $\theta_o(t),$ there holds
\begin{equation} \quad \int_0^{2\pi} \theta'_o(t)dt=0.\end{equation}
The last equation shows that the frequency function of an outer function is negative on a set of positive Lebesgue measures if the outer function itself is not identical to the zero function. Signals that have the property
\begin{equation} \theta'_\phi(t)+\theta'_s(t)+\theta'_o(t)\ge 0, \quad {\rm a.e.},\end{equation}
are called \emph{mono-components} \cite{Qianphasederivative}. A Hardy space function may not be a mono-component. However, any Hardy space function has a none-negative \emph{mean-frequency}, as defined in
\begin{Definition}
For a Hardy space function $f$ the quantity
\begin{equation} \int_0^{2\pi} \theta'_\phi(t)dt+\int_0^{2\pi} \theta_s'(t)dy\end{equation}
is called the \emph{mean-frequency} of $f,$ denoted as ${\rm M_F}(f).$
\end{Definition}
We note that ${\rm M_F}$ is additive in the sense ${\rm M_F}(fg)={\rm M_F}(f)+{\rm M_F}(g).$
\def\bN{\bf N}
\def\K{\mathcal K}
\def\tK{\tilde{\mathcal K}}
\def\tD{\tilde{\mathcal D}}

The phase derivative of a M\"obius transform
\[ \tau_a(z)=\frac{z-a}{1-\overline{a}z}\] is easily computed.
Let $\tau(e^{it})=e^{i\psi_a(t)}.$ Then $\psi_a'(t)$ is the Poisson kernel \cite{Garnett}
\[ P_{|a|}(e^{i(s-t)}), \quad a=|a|e^{is}.\]
Therefore,
\[ {\rm M_F}(\tau_a(z))=\frac{1}{2\pi}\int_0^{2\pi}P_{|a|}(e^{s-t})=1.\]

Due to the additivity, for $n,m\ge 0,$ ${\rm M_F}(z^n)=n,$ and
\[ {\rm M_F}(\prod_{k=1}^m \frac{z-a_k}{1-\overline{a}_kz})=m.\]
Since the Szeg\"o kernel $k_a$ and Szeg\"o dictionary elements $e_a,$ are outer functions,
\[ {\rm M_F}(k_a(z))={\rm M_F}(e_a(z))=0.\]
There exist standard models of functions in the Hardy space that form systems with increasing mean-frequencies. Let ${\bf a}=(a_1,\cdots,a_n)$ be an $n$-tuple of elements in the unit disc ${\bD}.$  Associated with ${\bf a}$ there exists an order-$n$ Takenaka-Malmquist (TM) system,
\[ \{B_k\}_{k=1}^n, \quad B_k(z)=e_k(z)\prod_{l=1}^{k-1}\frac{z-a_l}{1-\overline{a}_lz}.\]
$B_k$ is also denoted as $B_{a_1,\cdots,a_k}$ to specify the dependence on $a_1,\cdots,a_k.$
The $n$-tuple ${\bf a}$ can be extended to become an infinite sequence in ${\bD},$ and thus define an order-$\infty$ TM system $\{B_k\}_{k=1}^\infty.$ We use the notation $\{B_k\}$ to denote either a finite or an infinite TM system.  In this paper, unless otherwise specified, in an $n$-tuple or an infinite sequence of $\{a_k\},$ multiplicities of a number $a_k$ is allowed. A system $\{B_k\}$ is, although not necessarily complete, orthonormal in $H^2({\bD})=H^2(\partial {\bD})$ with the inner product given by (\ref{given}). It is known that an infinite TM system is a basis of
$H^2({\bD})$ if and only if the sequence ${\bf a}$ satisfies the hyperbolic non-separability condition
\[ \sum_{k=1}^\infty (1-|a_k|)=\infty.\]
The Fourier basis $\{z^k\}_{k=0}^\infty$ is a particular case corresponding to all $a_k\equiv 0, k=1,\cdots$

We have the following result.\\

\begin{thm}\label{frequency level}
 ${\rm M_F}(k_{n,a})={\rm M_F}(e_{n,a})=n,$ and ${\rm M_F}(B_n)=n-1.$
\end{thm}

\noindent \begin{pf} We note that $\frac{1}{(1-\overline{a}z)^{n+1}}$ is an outer function. As a consequence of the additivity, there holds
\[ {\rm M_F}(k_{n,a})=
{\rm M_F}(z^n)+{\rm M_F}\left(\frac{1}{(1-\overline{a}z)^{n+1}}\right)=n.\]
As a consequence of ${\rm M_F} (\tau_{a_l})=1$ and ${\rm M_F}(e_{a_n})=0,$ we have
\[ {\rm M_F}(B_n)={\rm M_F}\left(\prod_{l=1}^{n-1}\frac{z-a_l}{1-\overline{a}_lz}\right)=n-1.\]
\end{pf}

\begin{Definition}
A consecutive multiple Szeg\"o kernel $\tilde{k}_{a}$ with respect to the $m$-tuple $(a_1,\cdots,a_{m-1},a)$ is defined to be
\[ \tilde{k}_a(z)\triangleq k_{l(a)-1,a}(z),\]
where $l(a),$ with a little abuse of notation, denotes the repeating number of $a$ in $(a_1,\cdots,a_{m-1},a).$
In particular, if there is no repeating, that is $l(a)=1,$ then $\tilde{k}_a(z)=k_a(z).$
\end{Definition}

Computation gives
\[ k_{l(a)-1,a}=
(-1)^{l(a)-1}(l(a)-1)!\frac{z^{l(a)-1}}{(1-\overline{a}z)^{l(a)}}.\]

In \cite{QSAFD} the following result is proved.\\

\begin{thm}\label{GS}
Let $\{B_k\}$ be the TM system corresponding to any given finite or infinite sequence $(a_1,\cdots).$ Then,
up to unimodular multiplicative constants,
$\{B_k\}$ is the consecutive Gram-Schmidt orthonormalization of the consecutive multiple Szeg\"o kernel $\tilde{k}_{a_k}.$
\end{thm}
\section{Algorithms Adopting Matching Pursuit Methodology: Analysis and Comparison}\label{section:3}
Apart from $n$-Best approximation in Hilbert spaces with a BVC dictionary, there exist a number of iterative type algorithms adopting the matching pursuit methodology. These algorithms have common, as well as different features and individual effectiveness in terms of the reconstruction of signals. In this section, we review and analyze AFD, GA, OGA, and POAFD. With a criterion, we compare them and give them an order in terms of their reconstruction efficiencies. Among the four methods, AFD and POAFD  belong to the same class (the AFD type), and GA and OGA belong to another (the greedy type). The AFD type is based on delicate complex and harmonic analysis aiming at frequency decomposition and attainability of the supreme energy matching pursuit, while the greedy type is mostly in the general functional analysis context applicable to a Hilbert space with any dictionary. The AFD type is a generalization of Fourier theory, applicable for the reproducing kernel Hilbert space setting, having generalizations to multivariate functions (several complex variables and Clifford algebra variables) with vector- and matrix-values \cite{Q2D,ACQS1,ACQS2,QSW}. We below concentrate on the basic unit disc context.
\subsection{Adeptive Fourier Decomposition: AFD \cite{MQ2022}}
Let $(\H,\D)=(H^2({\D}),\{e_a\}_{a\in {\bD}}).$ It is proved in \cite{QW} that ${\D}$ is a dictionary in $H^2({\D})$ satisfying BVC.  For $f\in H^2({\D}), f_1=f,$ as a consequence of BVC, one can find
\[ a_1=\arg \max \{|\langle f_1,e_b\rangle|\ :\ b\in {\bD}\},\]
and
\[ \|f-\langle f_1,e_{a_1}\rangle e_{a_1}\|^2\] is hence minimized over all one-dimensional linear spaces generated by $e_b, b\in {\D}.$
Define Define, for $k \ge 2$,
\[ f_k(z)=\frac{f_{k-1}(z)-\langle f_{k-1},e_{a_{k-1}}\rangle e_{a_{k-1}}(z)}{\frac{z-a_{k-1}}{1-\overline{a}_{k-1}z}},\]
and select
\begin{eqnarray}\label{SL}
a_k=\arg \max \{|\langle f_k,e_b\rangle|\ :\ b\in {\bD}\}.\end{eqnarray}
 We call
$f_k$ the \emph{reduced remainder}, being the image of the $a_k$-\emph{generalized backward shift operator to} $f_{k-1}.$
It can be shown, inductively, $f_k\in H^2({\D})$, and
\begin{eqnarray}\label{AFD}
f(z)=\sum_{l=1}^k \langle f_l,e_{a_l}\rangle B_l(z)+f_{k+1}(z)\prod_{l=1}^k\frac{z-a_{l}}{1-\overline{a}_{l}z},\end{eqnarray}
where $\{B_l\}_{l=1}^k$ is the $k$-TM system defined by $(a_1,\cdots,a_k).$ We note that the above procedure allows multiplicity of the parameter $a_k$ selection. We note that the generalized backward shifts automatically generate the orthonormal $k$-TM system $\{B_l\}_{l=1}^k$. The orthogonality relations imply the useful relations
\begin{eqnarray}\label{useful}
\langle f_l,e_{a_l}\rangle=\langle f_l^\dag,B_l\rangle=\langle f,B_l\rangle,
\end{eqnarray}
where
\begin{eqnarray}\label{dagger} f^\dag_l\triangleq f-\sum_{j=1}^{l-1}\langle f_j,e_{a_j}\rangle B_j(z)=f_{l}(z)\prod_{j=1}^{l-1}\frac{z-a_{j}}{1-\overline{a}_{j}z}\end{eqnarray}
is the $l$-th-(AFD) \emph{orthogonal remainder}. We note that from the first identical relation in (\ref{dagger}) the $l$-th-(AFD) orthogonal remainder $f_{l}^\dag (z)$ is orthogonal with all the terms $B_j, j<l;$
and, from the second identical relation in (\ref{dagger}),
\[ \sum_{j=1}^{l-1}\langle f_j,e_{a_j}\rangle B_j(z)\]
is an interpolation rational function of $f$ at the points $a_1,\cdots,a_{l-1}.$ \\

It turns out that the partial sums converge to the original function $f$ \cite{QW}.
\[ f(z)=\sum_{l=1}^\infty \langle f_l,e_{a_l}\rangle B_l(z).\]
We will the convergence rates in the latter part of this section.

\subsection{Greedy Algorithm: GA \cite{MZ,Temlyakov,DT}}
Greedy algorithm is applicable to a general Hilbert space $\H$ with a dictionary $\D$. To compare it with other matching pursuit algorithms we assume that $\D$ satisfies BVC.

Let $f\in \H,$ and $g_1=f.$ The first matching pursuit step is the same as AFD:
Owing to BVC we can select
\begin{eqnarray} \label{as} q_1=\arg\max \{|\langle g_1,E_p\rangle|\ :\ p\in {\E}\}.\end{eqnarray}
There holds
\[ f=\langle f_1,E_{q_1}\rangle E_{q_1}+g_2.\]
Since $g_2$ is orthogonal with $E_{q_1},$ we have
\[ \|f\|^2=|\langle f_1,E_{q_1}\rangle|^2+\|g_2\|^2.\]
Due to the maximal selection of $q_1$ in (\ref{as}) the remaining energy in $g_2$ is minimized. Iteratively, define $g_k$ to be the $k$-th \emph{iterative remainder} given by
\begin{eqnarray}\label{SR}
f=\sum_{l=1}^{k-1}\langle g_l,E_{q_l}\rangle E_{q_l}+g_k,
\end{eqnarray}
where \begin{eqnarray}\label{in}q_l=\arg \max \{|\langle g_l,E_p\rangle|\ :\ p\in {\E}\}, l=1,\cdots,k-1.\end{eqnarray}
We note that for each $l=1,\cdots,k,$ the remainder $g_l$ is orthogonal with the last $E_{q_{l-1}}.$ There follows
\begin{eqnarray*}
\|f\|^2&=&|\langle g_1,E_{q_1}\rangle|^2+|\langle g_2,E_{q_2}\rangle|^2+\|g_3\|^2\\
&=&\sum_{l=1}^{k-1}|\langle g_l,E_{q_l}\rangle|^2+ \|g_k\|^2.
\end{eqnarray*}
The energy of $g_k$ decays to zero, and, as a consequence,
\begin{eqnarray}\label{GA}
f=\sum_{l=1}^{\infty}\langle g_k,E_{q_k}\rangle E_{q_k}.
\end{eqnarray}
\subsection{Orthogonal Greedy Algorithm: OGA \cite{MZ,Temlyakov,DT}}
We are under the same assumptions as for GA:
We have a pair $(\H,\D),$ where
$\D=\{E_q\}, q\in \E,$ is a dictionary of $\H$ satisfying BVC. Let $f\in \H$ and
$h_1=f.$ The first parameter selection is the same as that for AFD and GA: BVC implies that we are able to select
\[q_1=\arg\max \{|\langle h_1,E_p\rangle|\ :\ p\in {\E}\}.\]
There follows then
\[ h_2=f-{\rm Proj}_{\{E_{q_1}\}}(f)=g_2 \perp E_{q_1},\]
where we adopt the notation ${\rm Proj}_{\mathcal A}(f)$ for the orthogonal projection of $f$ into the linear span of the functions in ${\mathcal A}.$
We also use the notation $Q_{\mathcal A}\triangleq I-{\rm Proj}_{\mathcal A}={\rm Proj}_{\mathcal A^\perp}.$ $Q_{\mathcal A}$ is also called the \emph{Gram-Schmidt orthogonalization operator} (GS operator) \emph{with respect to the function set ${\mathcal A}.$}

We select
\begin{eqnarray}\label{superior}q_2=\arg\max \{ |\langle h_2,E_p\rangle|\ :\ p\in {\E}\}.\end{eqnarray}
Since $h_2=g_2,$ this maximal selection principle is the same as for GA.
Note that $E_{q_1}$ and $E_{q_2}$ are not necessarily orthogonal.
Define the $3^{rd}$ orthogonal remainder
\[h_3=f-{\rm Proj}_{\{E_{q_1},E_{q_2}\}}(f).\]
The Hilbert space property implies
\begin{eqnarray}\label{sup1}\|h_3\|\leq \|g_3\|.\end{eqnarray}
We will refer to this fact as $\lq\lq$OGA \emph{is }\emph{superior} \emph{to} GA", or say that as a matching pursuit algorithm $\lq\lq$OGA \emph{is} \emph{stronger} \emph{than} GA". Proceeding like this, we obtain the selections $\{E_{q_1},\cdots,E_{q_{k-1}}\}$ and formulate the $k$-th \emph{remainder} $h_k$ with the relation
\begin{eqnarray}\label{compared} f={\rm Proj}_{\{E_{q_1},\cdots,E_{q_{k-1}}\}}(f)+h_k,\end{eqnarray}
where
\begin{eqnarray} \label{no}q_l=\arg \max \{|\langle h_l,E_p\rangle|\ :\ p\in {\E}\}, l=1,\cdots,k-1.\end{eqnarray}
Consecutively, we select
\[ q_k=\arg \max \{|\langle h_k,E_p\rangle|\ :\ p\in {\E}\}.\]
It may be proved that $\|h_k\|\to 0$ as $k\to \infty.$  As a consequence, there holds
\[ f=\lim_{k\to \infty}{\rm Proj}_{\{E_{q_1},\cdots,E_{q_{k}}\}}(f)=\sum_{l=1}^\infty \langle f,E_{q_1,\cdots,q_{l}}\rangle E_{q_1,\cdots,q_{l}}
,\]
where $(E_{q_1},E_{q_1,q_2},\cdots,E_{q_1,\cdots,q_{l}})$ is the consecutive GS orthogonalization of
$(E_{q_1},{E}_{q_2},\cdots,{E}_{q_{l}}).$ We note that like GA
under OGA the selected parameters have no multiplicity.
\subsection{Pre-Orthogonal Adaptive Fourier Decomposition: POAFD}
As for GA and OGA we assume that
$\H$ is a general Hilbert space with a BVC dictionary
$\D=\{E_q\}, q\in \E.$ Since we will involve multiple kernels, we assume $E_q$ to be differentiable with respect to $q$ up to the needed orders. When $q$ is a vector, existence of directional derivatives is assumed. \\

The first step matching pursuit is again the same as that for AFD, GA, and OGA:
\[ q_1=\arg\max \{|\langle f,E_p\rangle|\ :\ p\in {\E}\}.\]
The selection of the second parameter $q_2$ now is different from that for GA and OGA (the latter two being the same at the $q_2$ selection):
\begin{eqnarray}\label{together} q_2=\arg\sup \{|\langle f,E_{q_1,q}\rangle|\ :\ q\in {\E}\},\end{eqnarray}
where $E_{q_1,q}$ is defined by the relation that $(E_{q_1},E_{q_1,q})$ is the GS orthonormalization of
$(E_{q_1},\tilde{E}_{q}).$ The function $E_{q_1,q}$ is unique up to a unimodular complex  multiplicative constant. Owing to this step the algorithm is called $\lq\lq$pre-orthogonal": The orthogonalization is done prior to the maximal selection.
In general, the POAFD maximal selection principle is
\begin{eqnarray}\label{to} q_n=\arg\sup \{|\langle f,E_{q_1,\cdots,q_{n-1},q}\rangle|\ :\ q\in {\E}\},\end{eqnarray}
where $(E_{q_1},E_{q_1,q_2},\cdots,E_{q_1,\cdots,q_{n-1}},E_{q_1,\cdots,q_{n-1},q}),$ regarded as TM system generated by $\D$ in $\H,$ is the consecutive GS orthogonalization of
$$(E_{q_1},\tilde{E}_{q_2},\cdots,\tilde{E}_{q_{n-1}},\tilde{E}q).$$

The $k$-th POAFD remainder denoted as $h_k\dag$, is defined through
\begin{eqnarray}\label{POAFD remainder} f=\sum_{l=1}^{k-1}\langle f,E_{q_1,\cdots,q_{l}}\rangle E_{q_1,\cdots,q_{l}}+h_k\dag,\end{eqnarray}
where the parameters are selected according to the POAFD maximal selection principle (\ref{to}). Both being orthogonal remainders, the $h_k^\dag$'s are different from the $h_k$'s for the latter are orthogonal remainder for OGA defined by (\ref{compared}) depending on the parameters selected according to OGA. POAFD was, in fact, suggested by the relation (\ref{useful}) in AFD in the Hardy space setting. In fact, writing  $Q_{\tilde{E}_{q_1},\cdots,\tilde{E}_{q_{l}}}$ briefly as $Q_{{q_1},\cdots,{q_{l}}},$ due to its properties as a projection, there holds
\[ \langle f,E_{q_1,\cdots,q_{l}}\rangle = \langle f,Q_{{q_1},\cdots,{q_{l-1}}}E_{q_1,\cdots,q_{l}}\rangle =\langle h_l\dag, E_{q_1,\cdots,q_{l}}\rangle,\]
which is (\ref{useful}) in which $h_l^\dag$ was written as $f_l^\dag$ to indicate its connection with the reduced AFD remainder $f_l.$
The coefficients in (\ref{POAFD remainder}) can also be written in the following forms
\begin{eqnarray} \label{relation} & &\left\langle  f,\frac{Q_{q_1,\cdots,q_{k-1}}(\tilde{E}_{q_k})}{\|Q_{q_1,\cdots,q_{k-1}}(\tilde{E}_{q_k})\|}\right\rangle \nonumber\\
&=&\left\langle Q_{q_1,\cdots,q_{k-1}}(f),\frac{Q_{q_1,\cdots,q_{k-1}}(\tilde{E}_{q_k})}
{\|Q_{q_1,\cdots,q_{k-1}}(\tilde{E}_{q_k})\|}\right\rangle
\nonumber\\
&=&\left\langle \frac{Q_{q_1,\cdots,q_{k-1}}(f)}{\|Q_{q_1,\cdots,q_{k-1}}(\tilde{E}_{q_k})\|},{E}_{q_k}\right\rangle.
\end{eqnarray}
As a reformulation of (\ref{useful}) this exhibits that, in a general Hilbert space with a BVC dictionary, POAFD is an analogous algorithm to Core AFD for the Hardy space. In such formation, in particular,
\[ \frac{Q_{q_1,\cdots,q_{k-1}}(f)}{\|Q_{q_1,\cdots,q_{k-1}}(\tilde{E}_{q_k})\|}\]
has the same role as the $k$-th \emph{reduced remainder}, being obtained in the Hardy space case through the generalized backward shift operators. Adopting the notation for the classical TM system,
\[B_k^{q_k}\triangleq \frac{Q_{q_1,\cdots,q_{k-1}}(\tilde{E}_{q_k})}
{\|Q_{q_1,\cdots,q_{k-1}}(\tilde{E}_{q_k})\|}\]
can be said from \emph{the TM system in the context}. And
\[Q_{q_1,\cdots,q_{k-1}}(f)\]
is the $k$-th \emph{orthogonal remainder} with respect to the POAFD selected parameters $q_1,\cdots,q_{k-1}.$
The relation (\ref{relation}) in particular, shows that the POAFD maximal selection principle corresponding to (\ref{to}) is performable.

For all the four types of matching pursuit algorithms, we have the following fundamental results \cite{MZ,Temlyakov,Q2D,DT,QWa}.\\

\begin{thm}\label{convergence}
Corresponding to the four different types (or contexts) of parameter maximal selection principles, namely (\ref{SL}), (\ref{in}), (\ref{no}), and (\ref{to}), the four remainders
$f_k^\dag, g_k, h_k$ and $h_k^\dag$ defined respectively through (\ref{dagger}), (\ref{SR}), (\ref{compared}), and (\ref{POAFD remainder}), all tend to zero in their respective Hilbert norms. Hence the corresponding partial sums all converge to the originally given signal $f.$ Moreover, if $f$ belongs to
\[H_M=\{ f\in \H\ :\ \exists \{q_k\}\subset \E, f=\sum_{k=1}^\infty c_kE_{q_k}, \sum_{k=1}^\infty|c_k| \leq M\},\]
then the norm of each of the above four types of $k$-remainders is dominated by $\frac{M}{\sqrt{k}}.$
\end{thm}

\begin{remark}
The rate $O(\frac{1}{\sqrt{n}})$ of convergence is valid for all matching pursuit algorithms \cite{DT}. The reference \cite{DT} constructs concrete examples to show that the convergence rate cannot be improved. Existence of such examples may also be asserted from the Karhunen-Loeve (KL) expansions. In fact, the KL expansion of the Brownian bridge, as an example, can be precisely estimated by
\[ {\mathbb E}\left[\|B-S_n\|^2_{L^2[0,1]}\right]=\sum_{j=n+1}^\infty \frac{1}{\pi^2j^2}\sim \frac{1}{\pi^2n}\]
(page 206 of \cite{LPS}). This estimation shows that there exist sample paths of Brownian bridge whose eigenfunction expansions have convergence rates as worse as $O(\frac{1}{\sqrt{n}}).$ On the other hand, due to the optimality of the KL expansion over all orthonormal expansions, there must exist matching pursuit expansions whose convergence rates are as worse as $O(\frac{1}{\sqrt{n}})$ as well.
\end{remark}

Efficiencies of the individual matching pursuit algorithms cannot be well compared in general. The parameters that give rise to the best matching pursuit may not be unique. The step by step optimality does not accumulate, and finally may not result in the overall optimality. Nevertheless, we can still draw a comparison under an intuitive criterion. We will first analyze the optimality of POAFD. \\

By using the GS operator, there holds, for $q\ne q_1,$
\[ E_{q_1,q}=\frac{{E}_q-\langle {E}_q,E_{q_1}\rangle E_{q_1}}{\|{E}_q-\langle {E}_q,E_{q_1}\rangle E_{q_1}\|}=
\frac{Q_{E_{q_1}}(E_q)}{\sqrt{1-|\langle E_q,E_{q_1}\rangle|^2}}.\]
%Since $Q_{E_{q_1}}$ is a projection, $Q^2_{E_{q_1}}=Q_{E_{q_1}}.$
Since $Q_{E_{q_1}}$ is self-adjoint,  there follows
\[ \langle f,E_{q_1,q}\rangle=\frac{\langle f,Q_{E_{q_1}}(E_q)\rangle}{\sqrt{1-|\langle E_q,E_{q_1}\rangle|^2}}=\frac{\langle Q_{E_{q_1}}f,E_q\rangle}{\sqrt{1-|\langle E_q,E_{q_1}\rangle|^2}}.\]
Replacing
$h_2=Q_{E_{q_1}}f,$ we have
\begin{eqnarray}\label{h}\sup_q |\langle f,E_{q_1,q}\rangle|&=&\sup_q |\langle h_2,E_{q_1,q}\rangle|\nonumber \\
&=& \sup_q |\frac{\langle h_2,E_q\rangle}{\sqrt{1-|\langle E_q,E_{q_1}\rangle|^2}}|\nonumber \\
&\ge& \sup_q |\langle h_2,E_q\rangle |.\end{eqnarray}
Due to the continuity, this argument is also valid for the limiting case $q=q_1.$
Recalling (\ref{together}), the last inequality shows that the POAFD energy matching is superior to that for OGA given in (\ref{superior}). As a consequence we have
\begin{eqnarray}\label{sup2} \|h_3^\dag\|\leq \|h_3\|.\end{eqnarray}

In the analyzed cases with the convention $f=f^\dag_1=g_1=h_1=h^\dag_1,$ we have $g_2=h_2, f^\dag_k=h^\dag_k, k\ge 3.$ It is observed that examples for which the strict inequality signs in (\ref{sup1}) or (\ref{sup2}) hold may be constructed. The above argument may be generalized: For $h_k$ being in the orthogonal complement of ${\rm span}\{B_1,\cdots,B_{k-1}\},$
\begin{align*} \langle f,B_k^q\rangle &= \langle h_k,\frac{Q_{q_1,\cdots,q_{k-1}}E_q}{\sqrt{1-\sum_{l=1}^{k-1}|\langle E_q,B_l\rangle|^2}}\rangle\\
 %&=\langle Q_{q_1,\cdots,q_{k-1}}g_k,\frac{E_q}{\sqrt{1-\sum_{l=1}^{k-1}|\langle E_q,B_l\rangle|^2}}\rangle\\
 &=\langle h_k,\frac{E_q}{\sqrt{1-\sum_{l=1}^{k-1}|\langle E_q,B_l\rangle|^2}}\rangle\\
 &=\frac{\langle h_k,E_q\rangle}{\sqrt{1-\sum_{l=1}^{k-1}|\langle E_q,B_l\rangle|^2}}.\end{align*}
In the next inequality-equality chain the left-end is the maximal selection principle of OGA, and right-end is for POAFD, showing that POAFD is superior to OGA.
\begin{align*}\label{right-end}
&\sup\{|\langle h_k,E_{q}\rangle| \ |\ q\in {\E}\}\\
&\leq
\sup\{\frac{|\langle h_k,E_{q}\rangle |}{\sqrt{1-\sum_{l=1}^{k-1}|\langle E_q,B_l\rangle|^2}}\ |\ q\in {\E}\} \\
&=\sup\{|\langle f,B^q_k\rangle |\ |\ q\in {\E}\}.
\end{align*}

Let, in general, Algorithm 1 and Algorithm 2 be among the concerned algorithms AFD, GA, OGA and POAFD, etc. If there exists a positive integer $k_0$ such that

\begin{enumerate}[(i):]
\item for any signal $f$ the energies of the $l$-remainders for $l<k_0$ of the two algorithms can be  made to be the same through their respective optimal matching pursuit selections; and
\item the energies of the $k_0$th remainder of Algorithm 1 are not larger than those of Algorithm 2, and for some particular signals $f$ strictly less  than those of Algorithm 2,
\end{enumerate}

then we say that Algorithm 1 is \emph{superior to} (\emph{or stronger than}) Algorithm 2.  In this case we write  Algorithm 1 $\ge$ Algorithm 2.  \\

\begin{thm}
For any Hilbert space with a dictionary satisfying BVC the associated algorithms satisfy
\[ POAFD\ge OGA\ge GA.\]
When the $n$-Best approximation exists, there holds
\[  n-{\rm Best}\ge {\rm any \ matching \ pursuit\ algorithm}.\]
\end{thm}
\section{Pre-Orthogonal Adaptive Fourier Decomposition with the Complete Szeg$\ddot{o}$ Dictionary}\label{section:4}
This section will be devoted to a detailed study of POAFD on the complete Szeg\"o dictionary.
Recall that in (\ref{kna}) and (\ref{ena}) we defined the multiple kernel $k_{n,a}$ and their normalizations, the dictionary elements $e_{n,a}.$ The Szeg\"o complete dictionary is denoted $\tD.$
The quantity of the norm of $k_{n,a}$ is computed as follows.\\

\begin{lem}\label{31}
\begin{align}
&\bigg\|\bigg(\frac{\partial}{\partial {\bar{a}}}\bigg)^{n}k_{a}(z)\bigg\|^{2}\nonumber\\
&=\sum_{m=0}^{n}C^{m}_{n}n!\frac{(2n-m)!}{(n-m)!}(a\bar{a})^{n-m}(1-\bar{a}a)^{m-2n-1}.
\end{align}
\end{lem}
\bigskip
\begin{pf}
\begin{align}
&\bigg\|\bigg(\frac{\partial}{\partial {\bar{a}}}\bigg)^{n}k_{a}(z)\bigg\|^{2}\nonumber\\
&=\bigg\langle\bigg(\frac{\partial}{\partial {\bar{a}}}\bigg)^{n}k_{a}(z),\bigg(\frac{\partial}{\partial {\bar{a}}}\bigg)^{n}k_{a}(z)\bigg\rangle\nonumber\\
&=\bigg(\frac{\partial}{\partial {a}}\bigg)^{n}\bigg\langle\bigg(\frac{\partial}{\partial {\bar{a}}}\bigg)^{n}k_{a}(z),k_{a}(z)\bigg\rangle\nonumber\\
&=\bigg(\frac{\partial}{\partial {a}}\bigg)^{n}\bigg(\frac{\partial}{\partial {\bar{a}}}\bigg)^{n}k_{a}(a).\nonumber
\end{align}
Through induction, we can show
\begin{equation*}
\bigg(\frac{\partial}{\partial {\bar{a}}}\bigg)^{n}k_{a}(a)=\frac{n!a^{n}}{(1-\bar{a}a)^{n+1}}.
\end{equation*}
Therefore,
\begin{align}\label{m=0}
&\bigg(\frac{\partial}{\partial {a}}\bigg)^{n}\bigg(\frac{\partial}{\partial {\bar{a}}}\bigg)^{n}k_{a}(a)\nonumber\\
&=\bigg(\frac{\partial}{\partial {a}}\bigg)^{n}\bigg(\frac{n!a^{n}}{(1-\bar{a}a)^{n+1}}\bigg)\nonumber\\
&=\sum_{m=0}^{n}C^{m}_{n}(n!a^{n})^{(m)}[(1-\bar{a}a)^{-n-1}]^{(n-m)}\nonumber\\
&=\sum_{m=0}^{n}C^{m}_{n}n!\frac{(2n-m)!}{(n-m)!}(a\bar{a})^{n-m}(1-\bar{a}a)^{m-2n-1}\nonumber\\
&= (2n)!\frac{|a|^{2n}}{(1-|a|^2)^{2n+1}}+\cdots .
\end{align}
\end{pf}

\begin{lem}\label{32}
Let $f\in H^{2}(\mathbb{D}).$ There holds for all $|a|<1$ uniformly
\begin{equation}\label{nbig}
\lim_{\hspace{0.1cm} n\rightarrow\infty} |\langle f,e_{n,a}\rangle|(z)= 0.
\end{equation}
\end{lem}

\begin{pf}
For any $\varepsilon>0$, due to $L^2$-convergence of Fourier series, there exists a polynomial $h(z)$ such that
\begin{equation*}
\|f-h\|<\varepsilon.
\end{equation*}
Denote by $N$ the  degree of the polynomial $h.$ When $n>N,$ in view of the Cauchy-Schwarz inequality and (\ref{view}), there holds for any $a\in {\bD}$
\begin{align}
|\langle f,e_{n,a}\rangle|&\leq |\langle f-h,e_{n,a}\rangle|+|\langle h,e_{n,a}\rangle|\nonumber\\
&<\varepsilon+|\langle h(z),e_{n,a}\rangle|\nonumber\\
&=\varepsilon+\bigg\langle h(z),\frac{\big(\frac{\partial}{\partial {\bar{a}}}\big)^{n}k_{a}(z)}{\|\big(\frac{\partial}{\partial {\bar{a}}}\big)^{n}k_{a}(z)\|}\bigg\rangle\nonumber\\
%&=\varepsilon+\frac{1}{\|\big(\frac{\partial}{\partial {\bar{a}}}\big)^{n}k_{a}(z)\|}\bigg\langle h(z),\bigg(\frac{\partial}{\partial {\bar{a}}}\bigg)^{n}k_{a}(z)\bigg\rangle\nonumber\\
&=\varepsilon+\frac{1}{\|\big(\frac{\partial}{\partial {\bar{a}}}\big)^{n}k_{a}(z)\|}\bigg\langle h(z)^{(n)},k_{a}(z)\bigg\rangle\nonumber\\
&=\varepsilon\notag.
\end{align}
\end{pf}

\begin{lem}\label{33}
For $f\in H^{2}(\mathbb{D})$ there holds uniformly for $n\in {\bN}$
\begin{equation}\label{BVCcomplete}
\lim_{|a|\rightarrow 1} |\langle f,e_{n,a}\rangle|= 0.
\end{equation}
\end{lem}

\begin{pf}
Given $\varepsilon>0.$ Due to Lemma~\ref{32}, we can restrict ourselves to verifying the convergence for $n\leq N.$
Since the span of the parametrized Szeg\"o kernels is dense in the whole space, the verification is further reduced to a
Szeg\"o kernel. For any but fixed $b\in {\bD},$
\begin{align}
&\bigg|\bigg\langle k_{b},\frac{\big(\frac{\partial}{\partial {\bar{a}}}\big)^{n}k_{a}}{\|\big(\frac{\partial}{\partial {\bar{a}}}\big)^{n}k_{a}\|}\bigg\rangle\bigg|\nonumber\\
&=\frac{1}{\|\big(\frac{\partial}{\partial {\bar{a}}}\big)^{n}k_{a}\|}\bigg\langle K_{b},\bigg(\frac{\partial}{\partial {\bar{a}}}\bigg)^{n}k_{a}\bigg\rangle\nonumber\\
&=\frac{1}{\|\big(\frac{\partial}{\partial {\bar{a}}}\big)^{n}k_{a}\|}\bigg|\bigg(\frac{\partial}{\partial {\bar{a}}}\bigg)^{n}k_{a}(b)\bigg|\nonumber\\
&\leq \frac{(1-|a|^{2})^{n+\frac{1}{2}}}{\sqrt{(2n)!}|a|^{n}}\frac{n!|b|^{n}}{|(1-\bar{a}b)|^{n+1}}\nonumber\\
&({\rm using \ the}\ m=0 \ {\rm \ term\ in\ (\ref{m=0})})\nonumber\\
&\leq\frac{(1-|a|^{2})^{n+\frac{1}{2}}}{\sqrt{(2n)!}|a|^{n}}\frac{n!|b|^{n}}{(1-|b|)^{n+1}}\rightarrow 0,\qquad {\rm as }\ |a|\to 1.\nonumber
\end{align}
\end{pf}

Lemma \ref{32} and Lemma \ref{33} together show that the complete dictionary satisfies BVC:

\begin{thm}\label{50}
For $f\in H^{2}(\mathbb{D})$ there holds
\begin{equation}\label{direct}
\lim_{|a|\rightarrow 1\hspace{0.1cm} or \hspace{0.1cm} n\rightarrow\infty} |\langle f,e_{n,a}\rangle|= 0.
\end{equation}
\end{thm}

In the last section, we deduced that in any Hilbert space with a BVC differentiate dictionary GA, OGA and POAFD can be performed. Owing to Theorem \ref{50}, taking $\H$ to be the Hardy space in the disc, and $\tilde{\D}=\{E_q\}, q\in \E,$ the complete Szeg\"o dictionary, where
\[ \E=\{q\in {\bN}\times {\bD}=\{(n,a)\ :\ n\in {\bN}, a\in {\bD}\}\]
and
\[ E_q=e_{n,a},\]
we conclude that GA, OGA, and POAFD can be performed with respect to the Szeg\"o complete dictionary.

The general theory of POAFD especially implies\\

\begin{thm}
For any $m-1$ previously given distinguished $2$-tuples $$(n_1,a_1),\cdots,(n_{m-1},a_{m-1}),$$ each in ${\bN}\times {\bD},$ there holds
\begin{equation}\label{BVCcomplete2}
\lim_{|a|\rightarrow 1\hspace{0.1cm} or \hspace{0.1cm} n\rightarrow\infty} |\langle f,B_m^{k_{n,a}}\rangle|= 0,
\end{equation}
where $B_m^{k_{n_m,a_m}}$ is the Hardy space function, unique up to unimodular multiple constants, characterized by the condition that $$(B_1^{k_{n_1,a_1}},\cdots,B_{m-1}^{k_{n_{m-1},a_{m-1}}},B_{m}^{k_{n_{m},a_{m}}})$$ is the GS orthonormalization of $(B_1^{k_{n_1,a_1}},\cdots,B_{m-1}^{k_{n_{m-1},a_{m-1}}},\tilde{k}_{n_{m},a_{m}}).$
We note that POAFD through the multiple kernel notion induces the completion of the dictionary in use. In our case, the dictionary in use by itself is a complete dictionary, whose completion, therefore, remains as just the same dictionary. As a consequence, there exists
$(n_m, a_m)\in {\bN}\times {\bD}$ such that
\begin{equation}\label{MSPcomplete} (n_m, a_m)=\arg \max \{ |\langle f,B_m^{k_{n,a}}\rangle|\ |\ (n,a)\in {\bN}\times {\bD}\}.\end{equation}
\end{thm}

The system $\{B_m^{k_{n,a}}\}$ has the role as the TM system made by the multiple Szeg\"o kernels, where they can be further specified by
\begin{align*} B_m^{k_{n,a}}&=
\frac{Q_{(n_1,a_1),\cdots,(n_{m-1},a_{m-1})}k_{n,a}}{\|Q_{(n_1,a_1),\cdots,(n_{m-1},a_{m-1})}k_{n,a}\|}\\
&=\frac{k_{n,a}-\sum_{l=1}^{m-1}\langle k_{n,a},B_l\rangle B_l}{\|k_{n,a}-\sum_{l=1}^{m-1}\langle k_{n,a},B_l\rangle B_l\|}\\
&=\frac{e_{n,a}-\sum_{l=1}^{m-1}\langle e_{n,a},B_l\rangle B_l}{\|e_{n,a}-\sum_{l=1}^{m-1}\langle e_{n,a},B_l\rangle B_l\|}.
\end{align*}
\section{Behaviour of Multiple Szeg\"o Kernel and Laguerre System in Hardy-Sobolev Space}\label{section:5}
For $\sigma>0,$ denoted by $H^{2}_{\sigma}({\bD})$ the order-$\sigma$ Hardy-Sobolev space:
\begin{eqnarray*}
 H^{2}_{\sigma}({\bD})
 =\{f(z)=\sum_{n=0}^\infty c_nz^n\ :
\ \sum_{n=0}^\infty |(1+n^\sigma)c_n|^2<\infty \}.\end{eqnarray*}
The space is briefly denoted $H^2_\sigma.$
For $f\in H^2_\sigma,$ define
$$\|f\|_{H^2_\sigma}\triangleq \left(\sum_{n=0}^\infty |(1+n^\sigma)c_n|^2\right)^{\frac{1}{2}}.$$
$H^2_\sigma$ is a subset of $H^2,$ being, by itself, a reproducing kernel Hilbert space. We will consider in this section expansions of $f\in H^2_\sigma$ by multiple Szeg\"o kernels in the norm of $H^2.$ By definition,  $f\in H^2_\sigma$ means that $f$ has up to the $\sigma$-th derivatives ($\sigma$ can be non-integer) of which all are of finite energy. Below we restrict ourselves to the integer $\sigma$ cases.\\

\begin{thm} If $f\in H^2_\sigma,$ where $\sigma$ is a non-negative integer, then
\begin{eqnarray}|\langle f,\frac{k_{n,a}}{\|k_{n,a}\|}\rangle|
\leq C(\sigma,a)\frac{1}{n^{\sigma}}\|f\|_{H^2_\sigma},\end{eqnarray}
where $C(\sigma,a)$ is a constant depending only on $\sigma$ and $a.$
\end{thm}

If $a=0,$ then $\frac{k_{n,a}}{\|k_{n,a}\|}=z^n,$ and the theorem reduces to a known classical result. In the general multiple kernel case, the proof uses estimates of the kernel function.

\begin{pf}

For $f\in H^{2}_{\sigma},$
\begin{align*}
&|\langle f,\frac{k_{n,a}}{\|k_{n,a}\|}\rangle|\\
&=\frac{1}{\|k_{n,a}\|}|\langle f,k_{n,a}\rangle|\\
&=\frac{1}{\|k_{n,a}\|}|\langle f^{(\sigma)},k_{n-\sigma,a}\rangle|\\
&\leq\frac{\|k_{n-\sigma,a}\|}{\|k_{n,a}\|}\|f\|_{H^{2}_{\sigma}}.
\end{align*}
By $\left(\frac{\partial}{\partial {\bar{a}}}\right)^{n}k_{a}(a)=\frac{n!a^n}{(1-\overline{a}a)^{n+1}}$, we get
\begin{align*}
&\frac{\|k_{n-\sigma,a}\|^{2}}{\|k_{n,a}\|^{2}}\\
&=\frac{k_{n-\sigma,a}(a)}{k_{n,a}(a)}\\
&=\frac{\left(\frac{\partial}{\partial {\bar{a}}}\right)^{2(n-\sigma)}k_{a}(a)}{\left(\frac{\partial}{\partial {\bar{a}}}\right)^{2n}k_{a}(a)} \\
&=\frac{\frac{[2(n-\sigma)]!a^{2(n-\sigma)}}{(1-\overline{a}a)^{2(n-\sigma)+1}}}{\frac{(2n)!a^{2n}}{(1-\overline{a}a)^{2n+1}}}\\
&=\frac{(2n-2\sigma)!}{(2n)!}\left(\frac{1-\overline{a}a}{a} \right)^{2\sigma}.
\end{align*}
By Stirling's formula,
\begin{align*}
&\frac{(2n-2\sigma)!}{(2n)!}\left(\frac{1-\overline{a}a}{a} \right)^{2\sigma}\\
&=\frac{\sqrt{2\pi(2n-2\sigma)}(\frac{2n-2\sigma}{e})^{2n-2\sigma}}{\sqrt{2\pi(2n)}(\frac{2n}{e})^{2n}}
\left(\frac{1-\overline{a}a}{a} \right)^{2\sigma}\\
&=\frac{1}{2^{2\sigma}}\frac{(n-\sigma)^{2n-2\sigma+\frac{1}{2}}}{n^{2n+\frac{1}{2}}}e^{2\sigma}
\left(\frac{1-\overline{a}a}{a} \right)^{2\sigma}\\
&\leq\frac{1}{2^{2\sigma}}e^{2\sigma}\left(\frac{1-\overline{a}a}{a} \right)^{2\sigma}3^{-\frac{\sigma}{n}(2n+\frac{1}{2})}\frac{1}{n^{2\sigma}}.
\end{align*}
Hence,
\begin{align*}
&|\langle f,\frac{k_{n,a}}{\|k_{n,a}\|}\rangle|\\
&\leq\frac{1}{2^{\sigma}}e^{\sigma}\left(\frac{1-\overline{a}a}{a} \right)^{\sigma}3^{-\frac{\sigma}{n}(n+\frac{1}{4})}\frac{1}{n^{\sigma}}\|f\|_{H^2_\sigma}
\end{align*}
tending to zero with the rate $\frac{1}{n^{\sigma}}$.
\end{pf}

When all the parameters $a_n$ are identical to $0$ the system of the normalized multiple kernels is a half of the Fourier orthonormal system.
When all $a_n$ are identical to some $a\ne 0,$ then the multiple kernels $k_{n,a}$ form a complete but un-orthogonal system. The GS orthonormalization of such multiple kernels forms the complete orthonormal Laguerre System
\[ B_{n,a}(z)=\frac{\sqrt{1-|a|^{2}}}{1-\bar{a}z}(\frac{z-a}{1-\bar{a}z})^{n-1}, \quad n=1,\cdots\]
(see Appendix of \cite{QSAFD}).
A Laguerre System is a particular TM system in which all $a_n$ are equal to a non-zero $a\in {\bD}.$

It is a classical result that Fourier series of functions in the $H^2_\sigma$ have convergence rate $O(n^{-\sigma}).$ This result is available for Laguerre Systems as well.\\

\begin{thm}\label{La}
For a function $f$ in the Sobolev space $H^2_\sigma$ there holds
\begin{equation}\|f-\sum^{\infty}_{l\le n}\langle f,B_{l,a}\rangle B_{l,a}\|\leq C_{a}\frac{1}{n^{\sigma}}\|f\|_{H^2_\sigma}.\end{equation}
\end{thm}

\begin{pf}
Performing change of variable $e^{i\theta}=\frac{e^{it}-a}{1-\bar{a}e^{it}},$ simple computation gives $dt=\frac{|1-\bar{a}e^{it}|^{2}}{1-|a|^{2}}d\theta$ and
\begin{align*}
&\langle f,B_{n,a}\rangle\notag\\
&=\frac{1}{2\pi}\int^{2\pi}_{0}f(e^{it})\frac{\sqrt{1-|a|^{2}}}{1-ae^{-it}}(\frac{e^{-it}-\bar{a}}{1-ae^{-it}})^{n-1}dt  \\
%&=\frac{1}{2\pi}\int^{2\pi}_{0}f(e^{it})\frac{\sqrt{1-|a|^{2}}}{1-ae^{-it}}\frac{|1-\bar{a}e^{it}|^{2}}{1-|a|^{2}}e^{-i(n-1)\theta}d\theta  \\
%&=\frac{1}{2\pi}\frac{1}{\sqrt{1-|a|^{2}}}\int^{2\pi}_{0}f(e^{it})\frac{|1-\bar{a}e^{it}|^{2}}{1-ae^{-it}}e^{-i(n-1)\theta}d\theta  \\
&=\frac{1}{2\pi}\frac{1}{\sqrt{1-|a|^{2}}}\int^{2\pi}_{0}f(e^{it})(1-\bar{a}e^{it})e^{-i(n-1)\theta}d\theta \\
&=\frac{1}{2\pi}\frac{1}{\sqrt{1-|a|^{2}}}\int^{2\pi}_{0}f(\frac{e^{i\theta}+a}{1+\bar{a}e^{i\theta}})(1-\bar{a}\frac{e^{i\theta}+a}{1+\bar{a}e^{i\theta}})e^{-i(n-1)\theta}d\theta \\
&=\frac{1}{2\pi}\frac{1}{\sqrt{1-|a|^{2}}}\int^{2\pi}_{0}F_a(\theta)e^{-i(n-1)\theta}d\theta\\
&=\frac{1}{\sqrt{1-|a|^{2}}}\langle F_a,e_{n,0}\rangle,
\end{align*}
where $F_a(\theta)=f(\frac{e^{i\theta}+a}{1+\bar{a}e^{i\theta}})
(1-\bar{a}\frac{e^{i\theta}+a}{1+\bar{a}e^{i\theta}}).$
 Since $|a|$ has a positive distance to $1,$ we have $\|F_a\|_{H^2_\sigma}\simeq \|f\|_{H^2_\sigma}.$
 Hence, through a brutal estimation,
\begin{align*}
&\|\sum^{\infty}_{l\geq n}\langle f,B_{l,a}\rangle B_{l,a}\|^{2}\\
&=\sum^{\infty}_{l\geq n}|\langle f,B_{l,a}\rangle |^{2}\\
&=\frac{1}{\sqrt{1-|a|^{2}}}\sum^{\infty}_{l\geq n}|\langle F_a,e_{n,0}\rangle |^{2}\\
&\leq \frac{1}{\sqrt{1-|a|^{2}}}\frac{1}{(1+n^\sigma)^2}\sum^{\infty}_{l\geq n}|(1+l^\sigma)\langle F_a,e_{n,0}\rangle |^{2}\\
&=\frac{1}{\sqrt{1-|a|^{2}}}\frac{1}{(1+n^\sigma)^2}\|F_a\|^2_{H^2_\sigma}\\
&\lessapprox C_{a}\frac{1}{(1+n^\sigma)^2}\|f\|^2_{H^2_\sigma}.
\end{align*}
\end{pf}

Below we provide the explicit transformation matrices between the $n$-Laguerre system and the corresponding first $n$ multiple kernels. \\

\begin{prop}\label{trans}
For arbitrary but fixed $n$ and $a\in \mathbb{D}$, $a\neq 0$, denote by the row matrix ${\mathcal B}=\{B_{l,a}\}_{l=1}^{n}$ the $n$-Laguerre system, and the row matrix ${\mathcal K}=\{k_{l,a}\}_{l=1}^{n}$ the corresponding $n$-tuple of multiple kernels. Then the invertible transformation matrix ${\mathcal T}$ such that ${\mathcal K}={\mathcal T}{\mathcal B}$ is given by ${\mathcal T}=\{c_{kj}\}_{n\times n},$
\begin{subnumcases}
{c_{kj}=}\nonumber
\frac{(j+1)!}{(k-j)!}\frac{a^{k-j}}{(1-|a|^{2})^{k+\frac{1}{2}}},& $j\le k\le n,$\\
\nonumber
0,& $n\ge j>k,$
\end{subnumcases}
 and ${\mathcal B}={\mathcal T}^{-1}{\mathcal K}$ with ${\mathcal T}^{-1}=\{d_{kj}\}_{n\times n},$
\begin{subnumcases}
{d_{kj}=}\nonumber
\frac{(-a)^{k-j}(1-|a|^{2})^{j-\frac{1}{2}}}{k!(k-j)!},& $j\le k\le n,$ \\
\nonumber
0,& $n\ge j>k.$
\end{subnumcases}

\begin{pf}
Since ${\rm span}\ {\mathcal B} = {\rm span}\ {\mathcal K}$ and
 $B_{l,a}$ are orthonormal, we have $k_{k,a}=\sum_{j=1}^{k}c_{kj} B_{j,a}$ , where
$c_{kj}=\langle k_{k,a},B_{j,a}\rangle.$ For $j\leq k\leq n,$
\begin{align*}
\overline{c}_{kj}&=\langle B_{j,a},k_{k,a}\rangle\\
&=\bigg\langle\frac{\sqrt{1-|a|^{2}}}{1-\bar{a}z}(\frac{z-a}{1-\bar{a}z})^{j-1},\frac{z^{k-1}}{(1-\bar{a}z)^{k}}\bigg\rangle\\
&=\frac{1}{2\pi}\int_{0}^{2\pi}\frac{\sqrt{1-|a|^{2}}}{1-\bar{a}e^{it}}(\frac{e^{it}-a}{1-\bar{a}e^{it}})^{j-1}\frac{e^{-i(k-1)t}}{(1-ae^{-it})^{k}}dt\\
&=\frac{1}{2\pi}\sqrt{1-|a|^{2}}\int_{0}^{2\pi}\frac{(e^{it}-a)^{j-1}}{(1-\bar{a}e^{it})^{j}}\frac{1}{(e^{it}-a)^{k}}de^{it}\\
&=\frac{1}{2\pi}\sqrt{1-|a|^{2}}\int_{\partial {\bD}}\frac{(\xi-a)^{j-1}}{(1-\bar{a}\xi)^{j}}\frac{1}{(\xi-a)^{k}}d\xi\\
&=\frac{1}{2\pi}\sqrt{1-|a|^{2}}\int_{\partial {\bD}}\frac{1}{(1-\bar{a}\xi)^{j}}\frac{1}{(\xi-a)^{k+1-j}}d\xi\\
&=\frac{j!}{(k-j)!}\frac{\bar{a}^{k-j}}{(1-|a|^{2})^{k-\frac{1}{2}}}.
\end{align*}
For $n\ge j > k$,
$B_{j,a}\bot {\rm span}\ {\mathcal B} = {\rm span}\ {\mathcal K}$,
and thus $B_{j,a} \bot k_{k,a}.$ We hence have $c_{kj} = 0$, $j > k.$

To compute the entries $d_{kj}$ of the inverse matrix ${\mathcal T}^{-1},$ by using mathematical induction, we have $d_{kj}=\frac{(-a)^{k-j}(1-|a|^{2})^{j-\frac{1}{2}}}{k!(k-j)!}, j\le k,$ and $d_{kj} = 0, j > k.$
\end{pf}
\end{prop}
\section{$n$-Best Approximation With the Complete Szeg\"o Kernel Dictionary: Existence and  Algorithm}\label{section:6}
Since derivatives of parameterized Szeg\"o kernels and multiple Szeg\"o kernels are still multiple Szeg\"o kernels,
the completion of the Szeg\"o complete dictionary $\{e_{m,a}\}, m=0,1,2,\cdots, a\in {\bD},$ is itself. Hence the $n$-Best problem with (\ref{diss}) returns to one with (\ref{dis}).\\

\begin{thm}
In the setting of the Hardy $H^2$ space with the complete Szeg\"o dictionary there exist solutions to the $n$-Best approximation problem.
\end{thm}

It is noted that the $n$-Best approximation with the complete Szeg\"o dictionary is different from that with the Szeg\"o kernel dictionary. For the former, the total number of the involved multiple kernels is required not exceeding $n,$ and for the latter, the sum of all the multiples of the involved parameters is required not exceeding $n.$  The theorem cannot be proved by directly invoking the existing $n$-Best results \cite{WQ3,QQLZ,Qian-n-best}. We outline a proof by referring to the one in \cite{WQ3}.

\begin{pf} Let $f$ be a fixed function in the Hardy space. We assume that $f$ is not identical to any $j$-Best approximation for $j<n,$ and hence exactly $n$ parameters are necessary to get an $n$-Best approximation. In this case, suppose that we have a sequence of $n$-tuples of parameter pairs $$\left((a_1^{(l)},m_1^{(l)}), \cdots, (a_n^{(l)},m_n^{(l)})\right), l=1,2,\cdots,$$ such that the corresponding $n$-tuples of multiple kernels $k_{a_j^{(l)},m_j^{(l)}}, j=1,\cdots,n,$ gives rise to, with the limit procedure $l$ $\to$ $\infty,$ the infimum of (\ref{dis}). We claim that the integers $m_j^{(l)}$ for $j=1,\cdots,n$ and $j=1,2,\cdots,$ have to be bounded, and the complex numbers $a_j^{(l)}$'s for $j=1,\cdots,n$ and $j=1,2,\cdots,$ have to be in a compact disc contained in ${\bD}.$ Once these claims are proved, by invoking the Bolzano-Weierstrass Theorem on existence of a convergent subsequence in the compact set we  conclude existence of $n$-Best approximation in our case.

The boundedness of $m_j^{(l)}$ is assured through the argument employed in the proof of Lemma \ref{32}. Next, by Proposition \ref{trans}, each of the involved multiple kernel $k_{a_j^{(l)},m_j^{(l)}}$ is a linear combination of the first $m_j^{(l)}$ functions of the corresponding Laguerre system. Then $(a_1^{(l)},m_1^{(l)}), \cdots, (a_n^{(l)},m_n^{(l)})$ altogether induce $n$ finite Laguerre systems of, respectively, the orders $m_1^{(l)}, \cdots, m_n^{(l)}.$ Put the $n$ finite Laguerre systems together and construct the equivalent orthonormal system by using the GS process. The obtained orthonormal system is a finite TM system.  The argument of the proof of the main result of \cite{WQ3} can be adopted to obtain that any $|a_j^{(l)}|$ tending to $1$ along with $l\to \infty$ will result in that the sequence $k_{a_j^{(l)},m_j^{(l)}}$ having no contribution in the approximation, and thus may be deleted. This concludes $n$ parameters are unnecessary. This is contradictory to the assumption at the beginning of the proof.
\end{pf}

\begin{remark}
Since the Fourier system is contained in the complete Szeg\"o dictionary, the $n$-Best approximation to  functions in $H^2_\sigma$ by the complete dictionary has a convergence rate at least as good as $O(n^{-\sigma}).$  Taking into account the convergence rete $O(n^{-1/2})$ for general matching pursuit algorithms, it is natural to guess, but we are unable to prove so far, that the $n$-Best approximation in $H^2_\sigma$ by the complete dictionary has the convergent rate $O(n^{-\sigma-1/2}).$
\end{remark}

We hereby cite two types of algorithms for finding one or all the $n$-Best solutions.
\subsection{A global and theoretical algorithm for finding all the $n$-Best solutions}
By a \emph{global method}, we refer to finding all the global minimum solutions, and, in particular, not being trapped in the local minimums. Such methods, therefore, have to be theoretical. Since, as asserted, the $m^{(l)}_j$ are altogether bounded, say, by $N,$ and all $a^{(l)}_j$ are in a compact disc $\overline{\bD}_r, 0<r<1,$ we reduce the problem to finding $\left(k_{m_1,a_1},\cdots,k_{m_n,a_n}\right)$ such that $(m_j,a_j)\in \{1,\cdots,N\}\times \overline{\bD}_r.$ This is to find all the global minimizers of the Lipschitz target function defined in the compact set. There exist theoretical, as well as practical methods to solve the type of problems in the optimization specialty. See, for instance, \cite{Qian-n-best,QDZC}, and the references thereby.
\subsection{A practical algorithm for finding an $n$-Best solution}
Practically, by the Gaussian gradient type or other similar methods based on local comparison of the target function values one may find an $n$-Best solution. By local comparison, although there is no guarantee of finding a global solution, it is practical and can often be used \cite{Bara,Bara2,Qian2,QWM}. We hereby recommend a local comparison type method, \emph{Cyclic POAFD}, with theoretical clarity and computational simplicity. It can be used with any dictionary \cite{Qian2}.

Suppose we have an initial $n$-tuple of dictionary elements $(e_{q^{(0)}_1},\cdots,e_{q^{(0)}_n}).$ The $n$-tuple of parameters $(q^{(0)}_1,\cdots,q^{(0)}_n)$ may be obtained randomly, or through an $n$-POAFD procedure. We rather use the kernels $k_{q^{(0)}_j}$ (pre-dictionary) instead of using the dictionary elements $e_{q^{(0)}_j}$ to save the computation of normalization. The idea is to improve the already obtained $n$-tuple of parameters in one by one cyclic manner: At each iteration step we replace a kernel already existing in the n tuple of kernels, and keep the other $n-1$ kernels unchanged. This cyclic procedure is performed continuously until each of the $n$-tuple of parameters cannot be improved with respect to a tolerant error threshold.
\section{Experiments}\label{section:7}
The experiments will select parameters in $\mathbf{D}$ which are expressed in polar coordinates:
\begin{equation}
a=\frac{x-1}{X}e^{i2\pi\frac{y-1}{Y}},
\end{equation}
for $x=1,2,\ldots,X$, and $y=1,2,\ldots,Y$, where $X$,
$Y$ correspond to the radial and angular discretization of the unit disc, respectively. We let $X=100$, $Y=100$ in the following experiments.
The abbreviations of the algorithms with the specifications including the type of dictionary in use, the type of matching pursuit algorithms, the iteration times, and whether it is $n$-Best etc. are given in Table \ref{tab0}. For Complete POAFD as an example, we obtain the partial sum series:
\begin{equation}
S_{K}=\sum^{K}_{k=1}\langle g_{k},B_{k}\rangle B_{k},
\end{equation}
with the relative error given by
\begin{equation}\label{relative error}
\frac{\|f-S_{K}\|}{\| f\|}.
\end{equation}
We take a similar definition to get the relative energy and relative error of other methods.
Below we use the following convention and notation: If for a particular signal $f$ the relative approximation error (\ref{relative error}) of Algorithm 1 is less than that of Algorithm 2, then we denote this fact by
\[ Algorithm 1 (f)\ge Algorithm 2 (f).\]

The simulations are performed in a MATLAB environment
using the data collected from the formula of the signal $f(z)$, $z=e^{it}$. The samples of $f$ are from $t_{j}=\frac{2\pi(j-1)}{100}, j=1,2,...,100.$
Six experiments are included.

\begin{table*}[!t]\scriptsize
\centering
\begin{tabular}{|c|c|}%通过添加 | 来表示是否需要绘制竖线
\hline %绘制横线
p-GA &  p iterations of GA algorithm with the Szeg$\ddot{o}$ dictionary\\
\hline
p-OGA &  p iterations of OGA algorithm with the Szeg$\ddot{o}$ dictionary\\
\hline
p-POAFD &  p iterations of POAFD algorithm with the Szeg$\ddot{o}$ dictionary\\
 \hline
p-Best &  p-Best approximation with the Szeg$\ddot{o}$ dictionary through some cycles\\
 \hline
p-Complete GA &  p iterations of GA algorithm with the complete Szeg$\ddot{o}$ dictionary\\
 \hline
p-Complete  OGA &  p iterations of OGA algorithm with the complete Szeg$\ddot{o}$ dictionary\\
  \hline
p-Complete POAFD &   p iterations of POAFD algorithm with the complete Szeg$\ddot{o}$ dictionary\\
\hline
p-Best Complete & p-Best approximation with the complete Szeg$\ddot{o}$ dictionary through some cycles\\
\hline
p-unwinding & p iterations of unwinding algorithm \\
\hline % 在表格最下方绘制横线
\end{tabular}
\caption{\label{tab0}}
\end{table*}

\begin{experiment}\label{ex1}
This experiment contains two lots of comparison of which one is 3 iterations of, respectively, GA, OGA, POAFD and n-Best algorithm, with respect to the Szeg\"o kernel dictionary; and the other is the same but with respect to the complete Szeg\"o dictionary. The tested toy function $f_{1}$ is given by the finite Blaschke form
\begin{equation}\label{53}
f_{1}(z)=\sum^{5}_{k=1}c_{k}B_{\{b_{1},...,b_{k}\}}(e^{it}),
\end{equation}
where $B_{\{b_{1},...,b_{k}\}}$ is the TM system
\begin{equation*}
B_{\{b_{1},...,b_{k}\}}=\frac{\sqrt{1-|a_{k}|^{2}}}{1-\bar{a}_{k}z}\prod^{k-1}_{l=1}\frac{z-a_{l}}{1-\bar{a}_{l}z},\hspace{0.2cm} a\in \mathbb{D},\hspace{0.2cm}k=1,2,...,
\end{equation*}
and $t\in (0,2\pi)$. The parameters $(b_{k}, c_{k})$, $k=1,2,\ldots,5$ are given in Table \ref{tab1}.

\begin{table}[!t]\scriptsize
\centering
\begin{tabular}{|c|c|c|}%通过添加 | 来表示是否需要绘制竖线
\hline % 在表格最上方绘制横线
\multicolumn{1}{|c}{Experiment \ref{ex1}} & \multicolumn{2}{c|}{}\\
\hline %绘制横线
 \multicolumn{1}{|c|}{$k$} & \multicolumn{1}{c|}{$b_{k}$} & \multicolumn{1}{c|}{$c_{k}$}\\
\hline %绘制横线
 1& -0.4750 + 0.3050i & -0.5861 - 0.04445i \\
 \hline
 2& -0.1800 + 0.7150i & 0.2428 - 0.6878i \\
 \hline
 3& 0.2600 - 0.7300i & 0.4423 - 0.3309i \\
 \hline
 4& 0.5400 + 0.3600i & -0.2703 - 0.8217i \\
 \hline
 5& -0.4850 - 0.2150i & -0.8085 + 0.3774i \\
\hline % 在表格最下方绘制横线
\end{tabular}
\caption{\label{tab1}}
\end{table}

\begin{figure}[!t]
\centering
\includegraphics[]{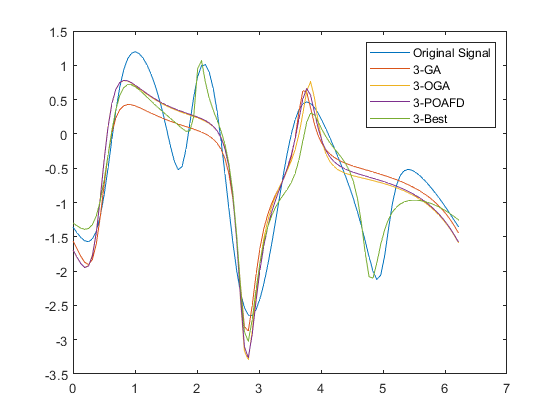}%
\caption{3 iterations with the Szeg$\ddot{o}$ dictionary.}
\label{fig1}
\end{figure}

\begin{table}[!t]\scriptsize%此处将表格字体设置为scriptsize，也可以设置为其它大小
\centering
\begin{tabular}{|c|c|c|c|c|}%通过添加 | 来表示是否需要绘制竖线
\hline % 在表格最上方绘制横线
\multicolumn{1}{|c}{Experiment \ref{ex1}} & \multicolumn{1}{c}{Fig.\ref{fig1}.} & \multicolumn{3}{c|}{}\\
\hline %绘制横线
3 iterations & GA & OGA & POAFD & 3-Best \\
\hline
relative error & 0.4581 & 0.4314 & 0.4246 & 0.2613 \\
\hline % 在表格最下方绘制横线
\end{tabular}
\caption{Relative error  \label{tab2}}
\end{table}

\begin{figure}[!t]
\centering
\includegraphics[]{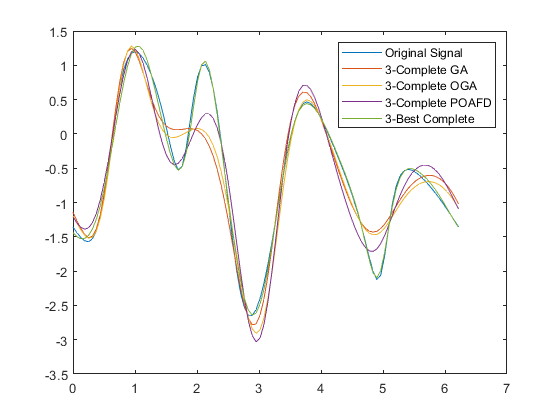}%
\caption{3 iterations with the complete Szeg$\ddot{o}$ dictionary.}
\label{fig2}
\end{figure}

\begin{table}[!t]\scriptsize%此处将表格字体设置为scriptsize，也可以设置为其它大小
\centering
\begin{tabular}{|c|c|c|c|c|}%通过添加 | 来表示是否需要绘制竖线
\hline % 在表格最上方绘制横线
\multicolumn{1}{|c}{Experiment \ref{ex1}} & \multicolumn{1}{c}{Fig.\ref{fig2}.} &\multicolumn{3}{c|}{}\\
\hline %绘制横线
3 iterations  & Complete GA & Complete OGA & Complete POAFD & 3-Best Complete\\
\hline
relative error  & 0.2941 & 0.2733 & 0.2590 & 0.0671\\
\hline % 在表格最下方绘制横线
\end{tabular}
\caption{Relative error  \label{tab3}}
\end{table}

We note that 3-Best by using POAFD with the Szeg$\ddot{o}$ dictionary to create the initial 3-tuple, and 3-Best Complete by using POAFD with the complete Szeg$\ddot{o}$ dictionary to create the initial 3-tuple.

According to the relative error given in Table \ref{tab2}, and results in Figure \ref{fig1}, we conclude that \emph{$n$-Best$(f_{1})$} $\geq$ \emph{POAFD$(f_{1})$} $\geq$ \emph{OGA$(f_{1})$} $\geq$ \emph{GA$(f_{1})$}.

From Table \ref{tab3}, and result in Figure \ref{fig2} we see that \emph{$n$-Best Complete$(f_{1})$} $\geq$ \emph{Complete POAFD$(f_{1})$} $\geq$ \emph{Complete OGA$(f_{1})$} $\geq$ \emph{Complete GA$(f_{1})$}. The order of superiority of the algorithms remains unchanged when using the complete dictionary, but by using the complete dictionary the approximation is more accurate.

Comparing POAFD and Complete GA algorithm, results in Figure \ref{fig3} and from the relative error given in Table \ref{tab2}, \ref{tab3} ,we conclude that \emph{Complete GA$(f_{1})$} $\geq$ \emph{POAFD$(f_{1})$}.

\begin{figure}[!t]
\centering
\includegraphics[]{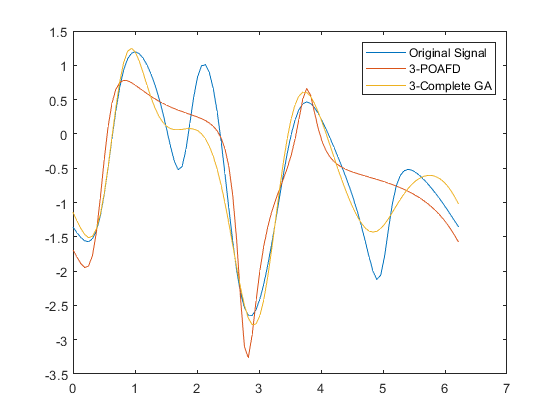}%
\caption{3 iterations of POAFD and Complete GA.}
\label{fig3}
\end{figure}
\end{experiment}

\begin{experiment}\label{ex3}
This experiment is to evaluate performs of POAFD with  the two dictionaries: the Szeg\"o one and the complete Szeg\"o one. The toy function is still (\ref{53}) but with the parameters $(b_{k}, c_{k})$, $k=1,2,\ldots,5$ given in Table \ref{tab3} that we note it $f_{2}$.

From Table \ref{tab6}, and results in Figures \ref{fig4}, \ref{fig5} we know that Complete POAFD can give a good approximation to $f$ through 9 iterations, while POAFD gives an approximation at a similar level through 18 iterations. Hence, \emph{Complete POAFD$(f_{2})$} $\geq$ \emph{POAFD$(f_{2})$}.

\begin{table}[!t]\scriptsize
\centering
\begin{tabular}{|c|c|c|}%通过添加 | 来表示是否需要绘制竖线
\hline % 在表格最上方绘制横线
\multicolumn{1}{|c}{Experiment \ref{ex3}} & \multicolumn{2}{c|}{}\\
\hline %绘制横线
 \multicolumn{1}{|c|}{$k$} & \multicolumn{1}{c|}{$b_{k}$} & \multicolumn{1}{c|}{$c_{k}$}\\
\hline %绘制横线
 1& -0.5850+0.2930i & -0.3861-0.0515i \\
 \hline %绘制横线
 2& 0.4806+0.2513i & -0.2802-0.7235i \\
 \hline %绘制横线
 3& 0.2505-0.6823i & 0.4505-0.4325i \\
 \hline %绘制横线
 4& -0.2005+0.6950i & 0.2539-0.7136i \\
 \hline %绘制横线
 5& -0.4512-0.1825i & -0.7562+0.4265i \\
\hline % 在表格最下方绘制横线
\end{tabular}
\caption{\label{tab5}}
\end{table}

\begin{figure}[!t]
\centering
\includegraphics[]{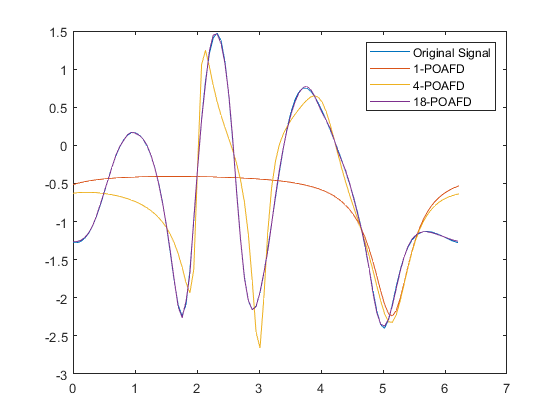}%
\caption{POAFD iterations with the Szeg$\ddot{o}$ dictionary.}
\label{fig4}
\end{figure}

\begin{figure}[!t]
\centering
\includegraphics[]{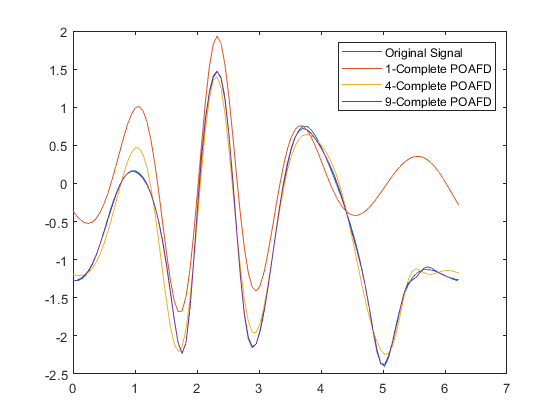}%
\caption{POAFD iterations with the complete Szeg$\ddot{o}$ dictionary.}
\label{fig5}
\end{figure}

\begin{table}[!t]\scriptsize%此处将表格字体设置为scriptsize，也可以设置为其它大小
\centering
\begin{tabular}{|c|c|c|c|c|c|c|}%通过添加 | 来表示是否需要绘制竖线
\hline % 在表格最上方绘制横线
\multicolumn{1}{|c}{Experiment \ref{ex3}} & \multicolumn{6}{c|}{}\\
\hline %绘制横线
Fig.\ref{fig4}. / Fig.\ref{fig5}. & \multicolumn{3}{c|}{POAFD} & \multicolumn{3}{c|}{Complete POAFD}\\
\hline %绘制横线
$K$ iterations & $K$=1 & $K$=4 & $K$=18 & $K$=1 & $K$=4 & $K$=9 \\
\hline %绘制横线
relative error  & 0.7906 & 0.4259 & 0.0173 & 0.7306 & 0.1334 & 0.0173 \\
\hline % 在表格最下方绘制横线
\end{tabular}
\caption{Relative error  \label{tab6}}
\end{table}
\end{experiment}

\begin{experiment}\label{ex4}
This experiment is to evaluate performs of $n$-Best with the two dictionaries: the Szeg\"o one and the complete Szeg\"o one. The toy function is still (\ref{53}) but with the parameters $(b_{k}, c_{k})$, $k=1,2,\ldots,4$ given in Table \ref{tab5} that we note it $f_{3}$.

We note that 8-Best by using POAFD with the Szeg$\ddot{o}$ dictionary to create the initial 8-tuple, and 4-Best Complete by using POAFD with the complete Szeg$\ddot{o}$ dictionary to create the initial 4-tuple.

From Table \ref{tab6}, and results in Figures \ref{fig6}, \ref{fig7} we know that $4$-Best Complete POAFD can give a good approximation to $f$ by 3 cycles, while $8$-Best POAFD gives an approximation at a similar level by 5 cycles. Hence, \emph{$n$-Best Complete$(f_{3})$} $\geq$ \emph{$n$-Best$(f_{3})$}.

\begin{table}[!t]\scriptsize
\centering
\begin{tabular}{|c|c|c|}%通过添加 | 来表示是否需要绘制竖线
\hline % 在表格最上方绘制横线
\multicolumn{1}{|c}{Experiment \ref{ex4}} & \multicolumn{2}{c|}{}\\
\hline %绘制横线
 \multicolumn{1}{|c|}{$k$} & \multicolumn{1}{c|}{$b_{k}$} & \multicolumn{1}{c|}{$c_{k}$}\\
\hline %绘制横线
 1& -0.4750+0.3050i & -0.5861-0.4444i \\
 \hline %绘制横线
 2& 0.3600-0.6300i & 0.4423-0.3308i \\
 \hline %绘制横线
 3& 0.5400+0.4600i & -0.2702-0.8217i \\
 \hline %绘制横线
 4& -0.4850-0.2150i & -0.7085+0.3773i \\
\hline % 在表格最下方绘制横线
\end{tabular}
\caption{\label{tab7}}
\end{table}

\begin{figure}[!t]
\centering
\includegraphics[]{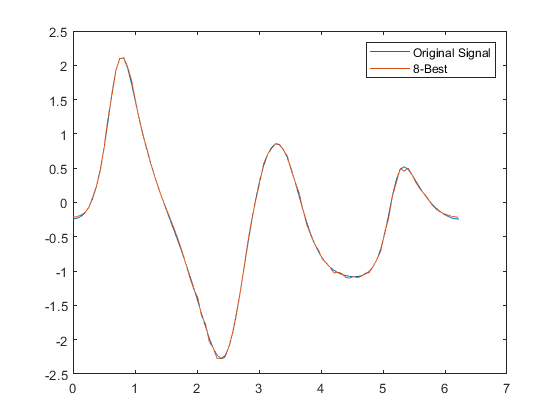}%
\caption{8-Best approximation with the Szeg$\ddot{o}$ dictionary through 5 cycles.}
\label{fig6}
\end{figure}

\begin{figure}[!t]
\centering
\includegraphics[]{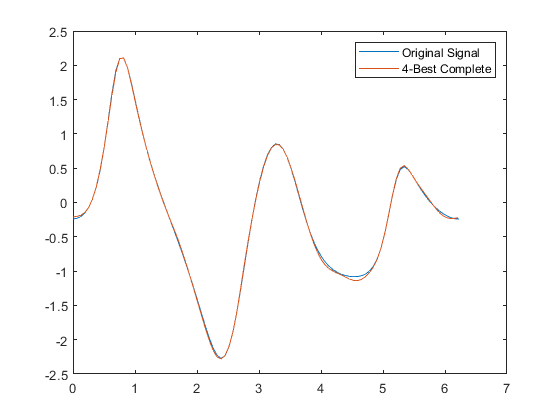}%
\caption{4-Best approximation with the complete Szeg$\ddot{o}$ dictionary through 3 cycles.}
\label{fig7}
\end{figure}

\begin{table}[!t]\scriptsize%此处将表格字体设置为scriptsize，也可以设置为其它大小
\centering
\begin{tabular}{|c|c|c|}%通过添加 | 来表示是否需要绘制竖线
\hline % 在表格最上方绘制横线
\multicolumn{1}{|c}{Experiment \ref{ex4}} & \multicolumn{2}{c|}{}\\
\hline %绘制横线
Fig.\ref{fig6}. / Fig.\ref{fig7}. & 8-Best POAFD & 4-Best Complete POAFD\\
\hline %绘制横线
$K$ iterations & $K$=8 & $K$=4\\
\hline %绘制横线
$l$ cycles &  $l$=5 &  $l$=3 \\
\hline %绘制横线
relative error   & 0.0224 & 0.0200 \\
\hline % 在表格最下方绘制横线
\end{tabular}
\caption{Relative error  \label{tab8}}
\end{table}
\end{experiment}

\begin{experiment}\label{ex5}
We note that unwinding Blaschke expansion is not a matching pursuit type algorithm, it, however, is of the same nature. As a very effective signal analysis method we add it to the pool of comparison \cite{QLS}.
This experiment is to compare unwinding, Complete POAFD, and $n$-Best algorithm, the signal is given by the samples of the following function
\begin{equation*}
f_{4}(z)=\frac{(z^{4}-d_{1})(d_{2}-z)^{5}}{(d_{3}-z)^{3}(d_{4}-z)^{2}},
\end{equation*}
where $t\in (0,2\pi)$. The parameters $d_{k}$,  $k=1,2,\ldots,4$ are given in Table \ref{tab9}.

We note that 2-Best Complete by using POAFD with the complete Szeg$\ddot{o}$ dictionary to create the initial 2-tuple.

From Table \ref{tab10}, and results in Figure \ref{fig8}, it is seen that \emph{$n$-Best Complete$(f_{4})$} $\geq$ \emph{Complete POAFD$(f_{4})$} $\geq$ \emph{unwinding$(f_{4})$}.

\begin{table}[!t]\scriptsize
\centering
\begin{tabular}{|c|c|}%通过添加 | 来表示是否需要绘制竖线
\hline % 在表格最上方绘制横线
\multicolumn{1}{|c}{Experiment \ref{ex5}} & \multicolumn{1}{c|}{}\\
\hline %绘制横线
 \multicolumn{1}{|c|}{$k$} & \multicolumn{1}{c|}{$d_{k}$} \\
\hline %绘制横线
 1& 3.1017-2.5305i \\
 \hline
 2& -6.1205+2.3674i \\
 \hline
 3& -5.4678-2.2502i \\
 \hline
 4& -4.4217+7.6913i \\
\hline % 在表格最下方绘制横线
\end{tabular}
\caption{\label{tab9}}
\end{table}

\begin{figure}[!t]
\centering
\includegraphics[]{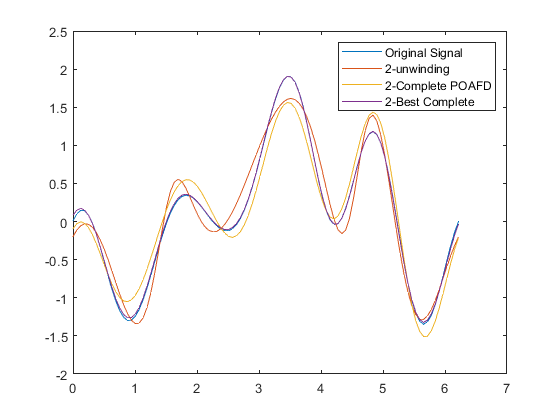}%
\caption{2 iterations of unwinding, Complete POAFD and $2$-Best Complete.}
\label{fig8}
\end{figure}

\begin{table}[!t]\scriptsize%此处将表格字体设置为scriptsize，也可以设置为其它大小
\centering
\begin{tabular}{|c|c|c|c|}%通过添加 | 来表示是否需要绘制竖线
\hline % 在表格最上方绘制横线
\multicolumn{1}{|c}{Experiment \ref{ex5}} & \multicolumn{1}{c}{Fig.\ref{fig8}.} & \multicolumn{2}{c|}{}\\
\hline %绘制横线
2 iterations &  unwinding & Complete POAFD & $2$-Best Complete \\
\hline %绘制横线
relative error & 0.2113  & 0.0863 & 0.0082\\
\hline % 在表格最下方绘制横线
\end{tabular}
\caption{Relative error  \label{tab10}}
\end{table}
\end{experiment}

\begin{experiment}\label{ex6}
Using Complete POAFD algorithm to denoise a noisy signal, the data are collected from the chirp signal without noise
\begin{equation*}
f(z)=e^{i\frac{t^{2}}{\pi}},
\end{equation*}
where $t\in (0,2\pi)$. And the original signal is a noisy signal with additive Gaussian white noise of $f$.

From the denoising effect given in Figure \ref{fig9}, we see Complete POAFD may denoise noisy signals.

\begin{figure}[!t]
\centering
\includegraphics[]{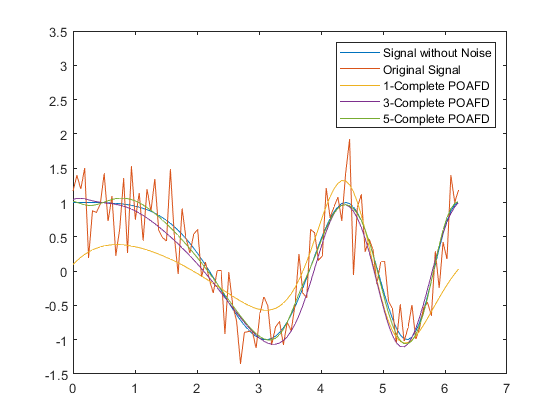}%
\caption{Denoising the chirp signal by Complete POAFD.}
\label{fig9}
\end{figure}
\end{experiment}

To summarize the experiments:
\begin{enumerate}[1.]
\item GA, OGA, POAFD (=AFD in HARDY), $n$-BEST with either the Szeg\"o or the complete Szeg\"o dictionary, and unwinding have the signal reconstruction efficiency following this order: \emph{$n$-Best Complete$(f_{1})$} $\geq$ \emph{Complete POAFD$(f_{1})$} $\geq$ \emph{Complete OGA$(f_{1})$} $\geq$ \emph{Complete GA$(f_{1})$} $\geq$ \emph{POAFD$(f_{1})$} $\geq$ \emph{OGA$(f_{1})$} $\geq$ \emph{GA$(f_{1})$}, \emph{$n$-Best Complete $(f_{3})$} $\geq$ \emph{$n$-Best$(f_{3})$}, and \emph{$n$-Best Complete$(f_{4})$} $\geq$ \emph{Complete POAFD$(f_{4})$} $\geq$ \emph{unwinding$(f_{4})$}.
\item In spite of the ordering theoretically or experimentally proved, referring to 1., experiments show that the differences between their efficiencies are not much.
\item Experiments show that \emph{Complete GA$(f_{1})$} $\geq$ \emph{POAFD$(f_{1})$}. A weaker algorithm with the complete dictionary is usually stronger than a stronger algorithm with a weaker dictionary, say the original Szeg\"o dictionary. This shows that it is the dictionary that is important.
\item Complete POAFD algorithm shows promising denoise effect.
\end{enumerate}
\section{Conclusion}\label{section:8}
The necessity of the notion multiple kernels is justified by existence of the natural question on $n$-Best kernel approximation, and by the desire for principal frequency decomposition of signals. For the latter, the concept mean-frequency is introduced. As the Fourier basis functions do, multiple kernels can extract the same rates of decaying to zero of signals in the Sobolev spaces. The work spells out the natural connections between the multiple Szeg\"o kernels and the Laguerre system. Well-posed-ness of $n$-Best approximation with the complete Szeg\"o dictionary is proved. With any dictionary, $n$-Best is the strongest, and POAFD is superior to all the other concerned matching pursuit methods in the one by one manner. Precisely, it is proved that the concerned matching pursuit methods from strong to weak are in the order n-Best, POAFD, OGA, and GA. Through concrete examples, we show that the complete Szeg\"o dictionary has great potential in sparse representation, for, especially, the weakest matching pursuit GA combined with the complete dictionary can inspire greater efficiency than what POAFD with the Szego dictionary does.

Future works include reducing the computational complexity and increasing the speed of the algorithm process.
\section*{Acknowledgments}
This work was supported by the research fund of Macau university of science and technology (grant no. FRG-22-075-MCMS).
The author wishes to express his sincere thankfulness to professor Tao Qian for his interest
and encouragement to study this topic, and useful comments on the draft material of this
article. The authors would like to thank the referees
for their helpful suggestions and comments.

\end{document}